\documentclass[showpacs,amsmath,amssymb,aps,twocolumn,longbibliography,pra]{revtex4-2}
\usepackage{graphicx}
\usepackage[colorlinks=true,citecolor=blue,linkcolor=magenta]{hyperref}
\usepackage[usenames]{color}
\usepackage{relsize}
\usepackage[english]{babel}
\usepackage{enumerate}
\usepackage{bbm}
\usepackage{xcolor}
\usepackage{framed}
\usepackage{bm}
\usepackage{braket}
\usepackage{algpseudocode,algorithm,algorithmicx}
\usepackage[most]{tcolorbox}
\tcbset{
  colback=blue!5!white,
  colframe=blue!75!black,
  fonttitle=\bfseries,
  coltitle=black,
  boxrule=0.8pt,
  arc=4pt,
  left=5pt,
  right=5pt,
  top=5pt,
  bottom=5pt,
}

\newcommand {\grsim} {\ {\raise-.5ex\hbox{$\buildrel>\over\sim$}}\ }
\newcommand {\lessim} {\ {\raise-.5ex\hbox{$\buildrel<\over\sim$}}\ }

\usepackage{xcolor}
\usepackage{siunitx}
\usepackage{booktabs}

\newcommand{\nocontentsline}[3]{}
\newcommand{\tocless}[2]{\bgroup\let\addcontentsline=\nocontentsline#1{#2}\egroup}

\newcommand{\RN}[1]{%
  \textup{\uppercase\expandafter{\romannumeral#1}}%
}

\usepackage{xr}
\begin{document}

\title{
Engineering multi-mode bosonic squeezed states using Monte-Carlo optimization
}

%

\author{Jieqiu Shao$^{1,2}$}
\author{Diego A. R. Dalvit$^{2}$ }
\author{Lukasz Cincio$^{2}$ }
\author{Bharath Hebbe Madhusudhana$^{3}$ }

\affiliation{$^{1}$Quantum New Mexico Institute, University of New Mexico, Albuquerque, NM-87107, United States}
\affiliation{$^{2}$ T-4, Los Alamos National Laboratory, Los Alamos, NM-87544, United States}
\affiliation{$^{3}$MPA-Quantum, Los Alamos National Laboratory, Los Alamos, NM-87544, United States}

\date{\today}
\begin{abstract}
Bosonic systems, such as photons and ultracold atoms, have played a central role in demonstrating quantum-enhanced sensing. Quantum entanglement, through squeezed and GHZ states, enables sensing beyond classical limits. However, such a quantum advantage has so far been confined to two-mode bosonic systems, as analogous multi-mode squeezed states are non-trivial to prepare. Here, we develop a Monte-Carlo based optimization technique which can be used to efficiently engineer a Hamiltonian control-sequence for multi-mode bosonic systems to prepare multi-mode squeezed states. Specifically, we consider a Bose-Einstein condensate in an optical lattice, relevant for applications in gravimetry and gradiometry, and demonstrate that metrologically useful squeezed states can be generated using the Bose-Hubbard Hamiltonian which includes on-site atomic interactions, tunable via Feshbach resonances. By analyzing the distribution (density) of the quantum Fisher information (QFI) over the Hilbert space, we identify a characteristic \textit{intermediate scaling} of the QFI: $\mathcal{O}(N^2 L+L^2 N)$, which lies between the standard quantum limit (SQL) and the Heisenberg limit (HL) for $N$ atoms in $L$ modes. We show that in general, within the Hilbert space there is a finite, $\mathcal{O}(1)$ measure subset of Hilbert space with an intermediate QFI scaling. Therefore, one can find a Hamiltonian control sequence  using a Monte Carlo optimization over random control sequences, that produces a state with intermediate scaling of the QFI.  We assume an experimentally accessible range of the control parameters in the Hamiltonian resources and show that the intermediate scaling can be readily achieved. Our results indicate that the HL can be approached in quantum gravimetry using realistic experimental parameters for systems with $L=\mathcal{O}(1)$ and $N\gg L$.
\end{abstract}

\maketitle

\section{Introduction} 
Atomic Bose-Einstein condensates (BEC) are a quintessential platform for quantum sensing~\cite{Montenegro_2025, Huang_2014}. They have been used in magnetometry~\cite{PhysRevLett.98.200801, GawlikHigbie2013} and gravimetry~\cite{PhysRevLett.67.181}. In all applications, spin-squeezing has been used to beat the standard quantum limit (SQL). Squeezing has been demonstrated through various methods, including the transfer of entanglement from non-classical states of light to atoms~\cite{PhysRevLett.83.1319}, scattering interactions between atoms~\cite{esteve2008squeezing, hamley2012spin}, and mediated interactions within a cavity~\cite{PhysRevLett.104.073602, hosten2016measurement}, achieving up to 20~dB of squeezing (i.e., a factor of 100 reduction in measurement variance). More recently, squeezed states have been prepared using Rydberg interactions between atoms in a tweezer array~\cite{eckner2023realizing}. Proof-of-principle quantum enhancement in Ramsey interferometry with atoms has also been achieved using squeezed states~\cite{PhysRevLett.109.253605, gross2010nonlinear}. However, most of the above demonstrations of squeezed states use two bosonic modes, where evolution under a constant Hamiltonian, e.g., one-axis twisting and two-axis counter-twisting can produce squeezing~\cite{PhysRevA.47.5138}. While squeezed states of multi-mode bosons have been studied theoretically~\cite{PhysRevA.89.032307}, experimentally viable techniques to produce them remain limited to three-component BECs~\cite{PhysRevA.102.043307, Hamley_2012, Huang_2017}. With larger number of modes, preparing squeezed states perhaps require more complicated quantum control, than evolution under a constant Hamiltonian. 

\begin{figure}[h!]
\centering
\includegraphics[width=0.96\linewidth]{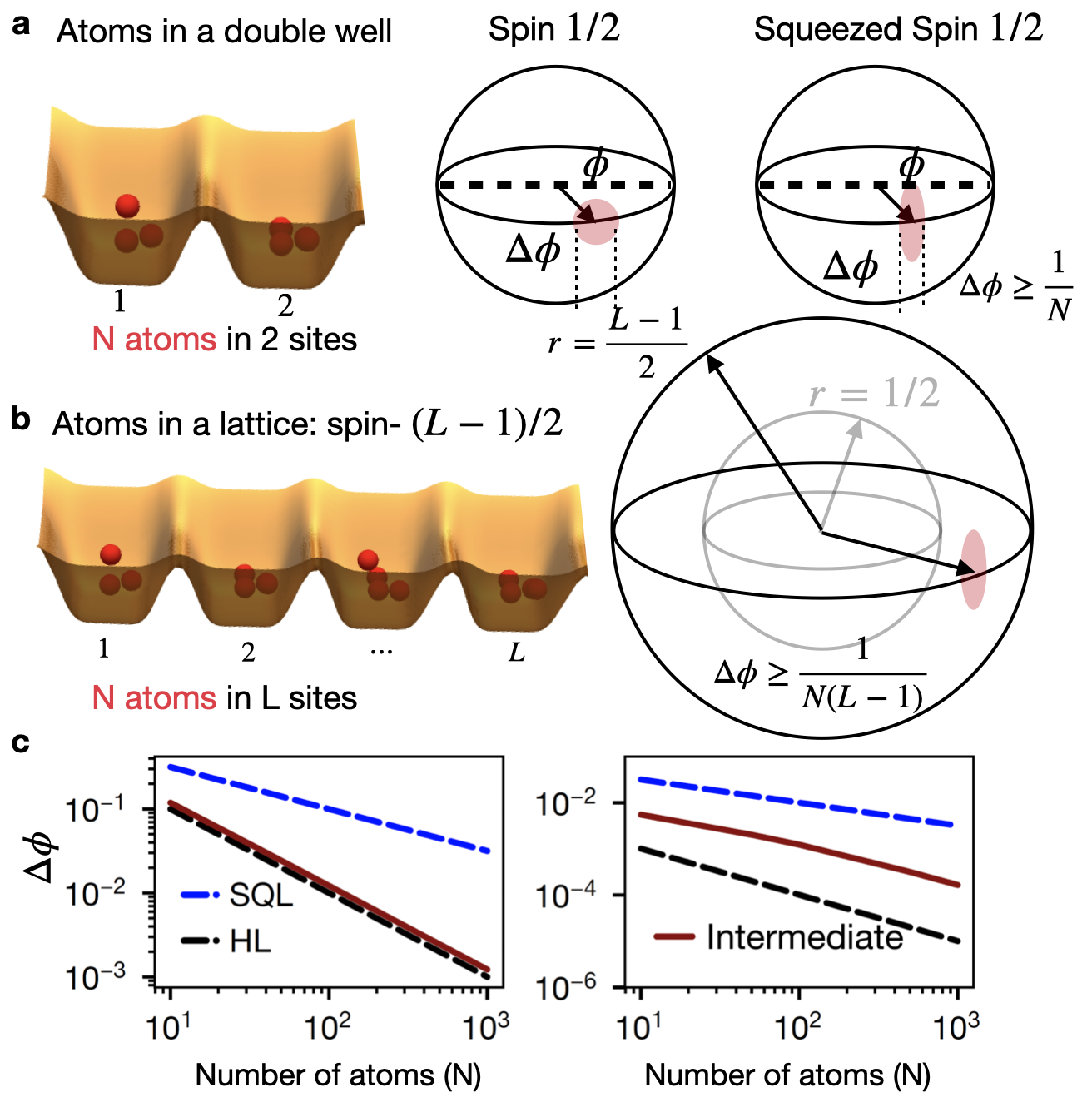}
\caption{\textbf{Squeezing in multi-mode bosons:} \textbf{a} Atoms in a double well can be used in quantum sensing and their states are represented by the standard Bloch sphere. \textbf{b} Bosons in a lattice with $L$ sites can be considered as a spin-$\frac{L-1}{2}$ system. This system can be used to measure gravity along the axis of the lattice (see text). The corresponding Heisenberg limit in quantum sensing with this system is $\frac{1}{N(L-1)}$.  \textbf{c.} Intermediate scaling (red line) between the SQL and the HL (see text). The left (right) panel shows the scaling for $L=2$ ($L=100$).  }
\label{Fig1}
\end{figure}

A possible solution is to use quantum optimal control (QOC)~\cite{Brif2010,Dong2010,Altafini2012,Q_Opt2015}, which is among the most popular methods to prepare quantum states with specific properties. In the context of model-based QOC, there are gradient-based methods like the GRadient Ascent Pulse Engineering algorithm (GRAPE) \cite{GRAPE,Khaneja_Brockett_2001} and Krotov methods \cite{Krotov1,Krotov2}, or Newton-based methods like Quantum PRojection Operator-based Newton method for Trajectory Optimization (Q-PRONTO)~\cite{Shao2022,Shao2024}. QOC has been used to produce optimized control sequences for squeezed states in Rydberg atom arrays~\cite{carrera2025preparingspinsqueezedstatesrydberg}. All of these techniques face a common challenge --- scalability. The Hilbert space dimension of multi-mode bosons scales exponentially both in the number of bosons and number of modes and therefore, QOC techniques tend to be inefficient, particularly when they require multiple evaluations of the time evolution and its gradients.

Here we show that a Monte Carlo–based technique can be used to efficiently optimize the control sequence for the Hamiltonian of multi-mode bosons. Considering a system of $N$ bosons in $L$ modes (Fig.~\ref{Fig1}a,b), we analytically show that our technique can be used to design sequences to prepare quantum states that surpass the SQL ($\mathcal{O}(NL^2)$) in terms of the QFI. We also show, however, that the technique approaches Heisenberg limit (HL, $\mathcal{O}(N^2L^2)$) only when the number of atoms is much larger than the number of modes, i.e., $N\gg L = \mathcal{O}(1)$—a condition satisfied in many important experimental platforms, including neutral atoms in optical double and triple wells and higher-spin BECs and is therefore experimentally very relevant. We also show a characteristic intermediate scaling, $\mathcal{O}(N^2L+L^2N)$, which lies between the SQL and HL which can always be reached using our Monte-Carlo technique (Fig.~\ref{Fig1}c). In contrast to the above mentioned deterministic QOC techniques, our Monte-Carlo method is a stochastic optimization technique that is especially effective at solving very complex, high-dimensional systems. Similar Monte-Carlo methods have been applied to quantum computing for sample problems~\cite{layden2023quantum}, quantum simulations~\cite{christmann2025}, and open quantum systems~\cite{Hryniuk2024tensornetworkbased}.


\section{System and problem setup}\label{sec:setup}
We consider an atomic BEC in an optical lattice, where each site can be occupied by multiple bosons (Fig.~\ref{Fig1}a,b). A gravitational field along the lattice introduces a ``tilt''. If the lattice is made sufficiently deep, the Hamiltonian of the system is dominated by this tilt:
\begin{equation}\label{Htilt}
    H_{\text{tilt}} = \eta \sum_{i=1}^L i \hat{a}_i^{\dagger} \hat{a}_i \ .
\end{equation}
Here, $\eta$ is a constant that depends on the strength of the gravitational field, $\hat{a}_i$ is the annihilation operator for site $i$, and $L$ is the number of lattice sites. Note the similarity of this operator to a spin operator
\begin{equation}
    S_z = \sum_{m=-S}^{S} m \hat{a}_m^{\dagger} \hat{a}_m,\quad S = \frac{L-1}{2} \ .
\end{equation}
Indeed, $H_{\text{tilt}}=\eta S_z-\frac{L+1}{2}$; the scalar $\frac{L+1}{2}$ has no practical relevance. In the limit of a shallow lattice, the tunneling part of the Hamiltonian dominates:
\begin{equation}\label{Htunnel}
    H_{\text{tunnel}} =  \sum_{i=1}^{L-1} J_{i, i+1} (\hat{a}_i^{\dagger} \hat{a}_{i+1} + \hat{a}_{i+1}^{\dagger} \hat{a}_i) \ .
\end{equation}
Here, $J_{i,i+1}$ are site-dependent parameters. While in a conventional optical lattice these hopping rates are typically equal to each other, i.e., $J_{i,i+1}=J$ and therefore cannot be tuned independently, recent advances in painted potentials~\cite{henderson2009experimental} can enable independent tuning of $J_{i,i+1}$. See sec.~\ref{Expt} for details.  Notice the similarity to the $S_x$ spin operator:
\begin{equation}
    S_x = \sum_{m=-S}^{S-1} \sqrt{(S-m)(S+m+1)} \left(\hat{a}_m^{\dagger} \hat{a}_{m+1} + \hat{a}_{m+1}^{\dagger} \hat{a}_m\right),
\end{equation}
and thus one can tune $J_{i,i+1}$ to simulate $S_x$.  A Ramsey sequence using these two Hamiltonians can be used to measure the phase $\phi =\eta \tau$ and therefore the parameter $\eta$. The corresponding precision is
\begin{equation}\label{eq:d_phi}
    \Delta\phi = \frac{1}{\sqrt{F}} \ ,
\end{equation}
where $F$ is the QFI (Eq.~\eqref{eq:GHZ}); the SQL and HL of this sensor with $N$ bosons are 
\begin{equation}
    \Delta \eta_{\text{SQL}} = \frac{1}{(L-1)\sqrt{N}\tau},\quad \Delta \eta_{\text{HL}} = \frac{1}{(L-1) N\tau} \ .
\end{equation}
Here, $\tau$ is the phase acquisition time, i.e., the time for which the system evolves under the Hamiltonian $H_{\text{tilt}}$. Finally, the on-site s-wave scattering interaction between atoms results in the term in the Hamiltonian:
\begin{equation}\label{Interaction}
    H_{\text{int}} = \sum_{i=1}^L U_i \hat{n}_i(\hat{n}_i-1) \ .
\end{equation}
Here, $\hat{n}_i = \hat{a}^{\dagger}_i\hat{a}_i$ is the number operator that counts the number of bosons; $U_i$ is the site-dependent strength, which again, cannot be controlled independently in conventional optical lattices. However, using more recent advances one can independently control the geometry of each trapping location and therefore control $U_i$ independently (see sec.~\ref{Expt} for details). This Hamiltonian can be used to entangle the atoms in the trap and thereby generate squeezing. For instance, in the simplest case of $L=2$ and uniform Hubbard interactions $U_i=U$, we directly obtain the one-axis twisting Hamiltonian:
\begin{equation}\label{sqz}
\begin{split}
    H_{\text{int}} = & U\left( \hat{n}_1^2 + \hat{n}_2^2 - (\hat{n}_1 + \hat{n}_2)\right)  = U \left( \frac{S_z^2 + \hat{N}^2}{2} - \hat{N}\right)\\
    = & H_{\text{squeezing}} \ .
\end{split}
\end{equation}
Note that $\hat{N}=\hat{n}_1 + \hat{n}_2$ is conserved and $S_z=\hat{n}_1 - \hat{n}_2$. For $L>2$, we show that Hamiltonian engineering can be used to develop a sequence that produces useful squeezing.

The figure-of-merit that quantifies the sensitivity to the tilt, i.e., gravity, is the QFI with respect to the operator $\hat{S}_z$. For a pure state $\ket{\psi}$ it is given by 
\begin{equation}\label{eq:QFI}
    F(\psi) = 4\left( \langle \hat{S}_z^2\rangle_{\psi}-\langle\hat{S}_z\rangle_{\psi}^2 \right) \ .
\end{equation}
Here, $\langle \cdot\rangle_{\psi}$ are the expectation values, e.g., $\langle S_z\rangle_{\psi} =\bra{\psi}S_z\ket{\psi}$. The QFI is maximized by the GHZ-like state
\begin{equation}\label{eq:GHZ}
    \ket{\psi_{\text{GHZ}}} = \frac{1}{\sqrt{2}}(\ket{N0 \dots 0} + e^{i\phi}\ket{0 \dots 0N}) ,
\end{equation}
where the Fock state $\ket{N0 \dots 0}$ denotes all $N$ bosons are in the first site, and $\ket{0 \dots 0N}$ denotes all $N$ bosons are in the $L-$th site. The QFI (Eq.~\eqref{eq:QFI}) of the above state (Eq.~\eqref{eq:GHZ}) is $F_\mathrm{HL}=N^2(L-1)^2$, which is the highest QFI one can obtain since the GHZ-like state saturates the HL. 

In most experimental realizations, Fock states $\ket{N 0\ldots0}$ and $\ket{N/L,N/L,\ldots, N/L}$ can be initialized. Also, coherent states, i.e., $\sum_{n_1 \ldots n_L} \sqrt{\frac{N!}{n_1!\cdots n_L!}}\ket{n_1\ldots n_L}$ can be initialized using non-interacting dynamics. The state $\ket{\psi}$, to be used for quantum sensing, can be prepared by evolving the Fock state or the coherent state under a time-dependent, controllable, interacting Hubbard Hamiltonian described above. We assume that the initial state is prepared during a time window $[0, T]$, before phase acquisition.  The time-dependent Hamiltonian can be engineered to prepare a desired state during this time window. In addition to hopping and Hubbard interactions, a controllable on-site potential $\Delta^{\text{ctrl.}}_i$ can be applied to each site. The full Hamiltonian of the system is,
\begin{equation}\label{eq:H}
\begin{split}
H = &\sum_{i=1}^{L-1} J_{i,i+1}
    \left( \hat{a}_i^{\dagger}\hat{a}_{i+1} + \hat{a}_{i+1}^{\dagger}\hat{a}_i \right) \\
  &+ \sum_{i=1}^{L} U_i \, \hat{n}_i (\hat{n}_i - 1) \\
  &+ \sum_{i=1}^{L} \left( \Delta_i^{\mathrm{ctrl.}} + \eta \, i \right) \hat{n}_i \ .
\end{split}
\end{equation}
 We assume that each parameter, $J_{i, i+1}$, $U_i$ and $\Delta^{\text{ctrl.}}_i$ can be tuned independently in time during state preparation. Moreover, the parameter $\eta$ typically has an unknown part which we intend to measure. For instance, in applications of gravimetry to fundamental physics, $\eta$ has a known contribution from gravity and a much weaker, unknown contribution from e.g., a fifth force, dark matter or dark energy, which we intend to measure. The known part of $\eta$ can be canceled by tuning $\Delta_i^{\text{ctrl.}}$ appropriately. We also assume that the unknown part is sufficiently small compared to the controllable $\Delta^{\text{ctrl.}}_i$, so that it can be ignored during state preparation. It becomes relevant during phase acquisition. Hereafter, we set $\Delta_i=\Delta_i^{\text{ctrl.}}+\eta i$ and assume $\Delta_i$ can be tuned arbitrarily.  We use QOC to optimize the QFI of the state $\ket{\psi}$ produced after evolving a Fock state (or the coherent state). We develop some background on the properties of QFI before optimizing it.



\section{Scaling and density of QFI}\label{sec:scaling}
We will discuss some useful properties of the QFI for this system. Given the two parameters $N$ and $L$, there are multiple regimes of interest in relation to the QFI. We consider two regimes:  (i) $N\gg1, \  L=\mathcal{O}(1)$ and (ii) $N, L\gg 1$. The dimension of the Hilbert space is 
\begin{equation}\label{eq:dim}
    d=\binom{N+L-1}{N} \ .
\end{equation}
While the exact Heisenberg limit is $F_\mathrm{HL}=N^2(L-1)^2$, we represent its order --- $N^2L^2$.  

To understand the QFI, we will consider its distribution $P(F)$ over the Hilbert space under Haar-random states. The mean value $(\mu)$ and the standard deviation ($\sigma$), to the highest order are given by
\begin{equation}\label{mean_std}
\begin{split}
    \mu = \frac{N^2L+NL^2}{3} \\
    \sigma = \frac{N^2L + NL^2}{3\sqrt{d/2}}
\end{split} 
\end{equation}
Here, $\mu = \int F(\psi) d\psi$ and $\int F^2(\psi)d\psi = \sigma^2 + \mu^2$. See sec.~\ref{sec:mean_std} of the supplementary information~\cite{supplements} for exact expressions and derivations.  For a desired QFI $F_0$, we consider the probability that a random state has a QFI higher than $F_0$, i.e., $p(F_0)=1-\int _0^{F_0} P(F)dF$. Here, $P(F)$ is the probability distribution of $F$ over the Hilbert space as defined earlier.  The value of $p(F_0)$ determines the \textit{density} of states with QFI larger than $F_0$ and that determines how hard is it to find a state (preparation) with QFI larger than $F_0$. We invoke the one–sided Chebyshev (Cantelli) inequality to estimate $p(F_0)$. 
\begin{equation}\label{chebyshev}
\begin{split}
    p(F_0)\leq  \frac{\sigma^2}{\sigma^2 + (F_0-\mu)^2} \text{\quad when } F_0\geq \mu\\ p(F_0) \geq \frac{(\mu-F_0)^2}{\sigma^2 + (\mu-F_0)^2} \text{\quad when } F_0\leq \mu
\end{split}
\end{equation}
See sec.~\ref{sec:chebyshev} of the supplementary information~\cite{supplements} for details and derivation. In particular, $p(F_0=\mu-\sigma)\geq 1/2$. That is, more than half of the states (by measure) have a QFI higher than $\mu-\sigma$, making the latter a characteristic scale of QFI in the system. We refer to the scaling corresponding to $\mu-\sigma$ as the \textit{intermediate scaling} which lies between the Heisenberg limit and Standard Quantum Limit:
\begin{equation}\label{intermediate_scaling}
\begin{split}
   \mathcal{O}( NL^2)&\leq \mathcal{O}({NL^2+N^2L})\leq \mathcal{O}(N^2L^2)\\
   \text{SQL}& \quad\quad \text{Intermediate}\quad\quad \ \text{HL}
\end{split}
\end{equation}
The intermediate scale represents the QFI of a \textit{typical state} in the Hilbert space.

Therefore, for the general case $N, L\gg 1$, one can obtain the intermediate scaling using a Monte-Carlo optimization. We will use these properties of the QFI to show that there exists an efficient Monte-Carlo optimization technique that can always find a control parameter trajectory which results in the intermediate scaling of the QFI. 

\section{Monte-Carlo optimization}
\begin{figure}
\centering
\includegraphics[width=0.95\linewidth]{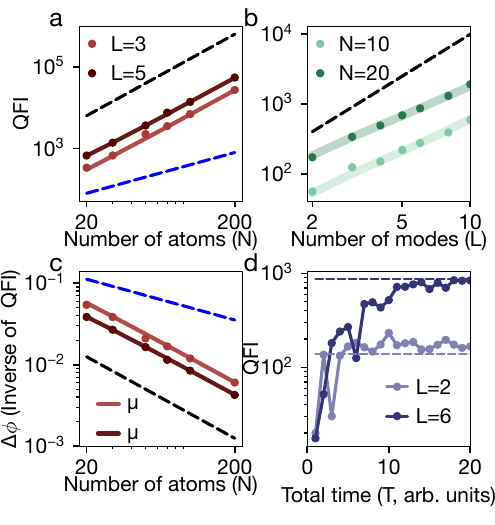}
\caption{\textbf{Monte-Carlo optimization:} \textbf{a.} QFI optimized using Monte-Carlo for $L= 3, 5$ and $N$ up to $200$. The markers correspond to the numerically optimized QFI using $\nu=10$ samples. The solid lines are the intermediate scaling, corresponding to $\mu$ in Eq.~(\ref{mean_std}). \textbf{b.} Similar plot for various $L$. \textbf{c.} The expected error $\Delta \phi \sim 1/\sqrt{F}$ as a function of $N$ (data from part (a)). \textbf{d.} The optimal QFI obtained using Monte-Carlo, with $N=20$, $\nu=10$ for various evolution times $T$. The dashed lines show the intermediate scaling $\mu$ for the same cases. The states $\psi(T)$ are not sufficiently entangled for small $T$ resulting in this asymptotic behavior. The code used to produce the above data is described in sec.~\ref{sec:code_description} of the supplementary information~\cite{supplements}.}
\label{Fig_monte_carlo}
\end{figure}

Eq.~(\ref{chebyshev}) with $F_0=\mu-\sigma$, i.e, the intermediate scaling, has another interesting consequence: among $\nu$ random states, at least one of them will have a QFI higher than $\mu-\sigma$ with a probability $1-(1-p(F_0))^{\nu} \geq 1-1/2^{\nu}$. In other words, if we pick $\nu$ random control trajectories and compute the corresponding QFIs of the final state: $F_1, \cdots, F_{\nu}$ and maximize, i.e., find $r_{\max}=\text{argmax}\{F_r\}$, then $F_{r_{\max}}\geq \mu-\sigma$ with a very high probability ($1-1/2^{\nu}$). Therefore, $r_{\max}$ very likely corresponds to a control trajectory that produces a quantum state with an intermediate scaling of QFI, if the random control trajectories generate \textit{Haar random states}. The strength of the control parameters and the total time $T$ taken for the state preparation are crucial here --- they determine if the final state is Haar random. We explore this further in Fig.~\ref{Fig_monte_carlo}d.

We refer to this technique --- of picking $\nu$ random trajectories and maximizing the QFI of the corresponding final states --- as the Monte-Carlo optimization. We demonstrate the Monte-Carlo optimization for a few examples with $L=\mathcal{O}(1)$ and $N\gg L$. For each random trajectory in the  Monte-Carlo optimization, we start with the same initial state, $\ket{\psi(0)}=(\hat{a}_1^{\dagger})^N/\sqrt{N!}\ket{\text{vac}}$, evolve it up to time $T$ (here, $T$ is the time taken for the state preparation) and compute the QFI, $F(\psi(T))$. The state-preparation time-interval $[0, T]$ is divided into $3n$ uniform sections for an integer $n$, where the size of each section is $\Delta t = T/(3n)$. Note that the Hamiltonian is time-independent within each $\Delta t$. Further, we evolve the system under the three Hamiltonians below sequentially. At the $k$-th step,

\begin{eqnarray}\label{H1H2H3}
    \begin{split}
        H_1^{(k)}&=\sum_i \Delta_i^{(k)} \hat{n}_i + U_i \hat{n}_i(\hat{n}_i-1)\\
        H_2^{(k)}&=\sum_i J_{2i}^{(k)} (\hat{a}^{\dagger}_{2i+1}\hat{a}_{2i}+\hat{a}^{\dagger}_{2i}\hat{a}_{2i+1})\\
        H_3^{(k)}&=\sum_i J_{2i-1}^{(k)} (\hat{a}^{\dagger}_{2i}\hat{a}_{2i-1}+\hat{a}^{\dagger}_{2i-1}\hat{a}_{2i})
    \end{split}
\end{eqnarray}
The final state is 
\begin{equation}\label{eq:final}
\ket{\psi(T)} = \Pi_ke^{-i\Delta t H_3^{(k)}}e^{-i\Delta t H_2^{(k)}}e^{-i\Delta t H_1^{(k)}}\ket{\psi(0)} \ . 
\end{equation}
In this picture, $H_1$ is diagonal in the number basis and $H_2, H_3$ can be relatively efficiently exponentiated. Each trajectory corresponds to $3$ random matrices --- a $n\times (L-1) $ matrix $J^{(k)}_i$ (it can be treated as a matrix because $k=1,\cdots, n$ and $i=1, \cdots, L-1$), and $n\times L$ matrices $\Delta^{(k)}_i$ and $U^{(k)}_i$. We compute the QFI at the final time for $\nu$ such random trajectories and obtain the maximum. 
In each random trajectory, we pick a matrix $J^{(k)}_i$ with uniformly random entries from $[0,1]$, and matrices $\Delta^{(k)}_i$ and $U^{(k)}_i$ with uniformly random entries from $[-1,1]$. We then evolve the initial state with timestep $\Delta t$ up to time $T$. 



        




\begin{algorithm}[H]
\caption{~~Monte-Carlo Optimization}\label{alg:MC_op}
\begin{algorithmic}[1]
  \vspace{5 pt}\Statex \begin{center}
          \textbf{Initialization}\end{center}
      \State Number of sites $L$, particles $N$, evolution time $T$, number of random trajectories $\nu$, and time steps $n$
          \Statex \emph{Initial state}
          \State $\ket{\psi(0)} = \frac{(\hat{a}_1^{\dagger})^N}{\sqrt{N!}}\ket{\text{vac}}$
          \Statex \emph{Divide the total time interval into $3n$ uniform sections}
          \State $\Delta t = T/(3n)$

      \Statex \begin{center}
          \textbf{Main Loop}
      \end{center}
      \For{$r = 1, \dots, \nu$} 
          \Statex \emph{Generate random matrices}
      \State $J^{(k)}_i \in [0,1],\qquad i=1,\dots,L-1,\quad k = 1,\dots,n$
      \State $\Delta^{(k)}_i \in [-1,1],\quad i=1,\dots,L,\quad k = 1,\dots,n$
      \State $U^{(k)}_i \in [-1,1],\quad i=1,\dots,L,\quad k = 1,\dots,n$
      \Statex \emph{Propagate the state Eq.~\eqref{eq:final}}
      \State $\ket{\psi_r(T)} = \prod_k e^{-i\Delta t H_3^{(k)}} e^{-i\Delta t H_2^{(k)}} e^{-i\Delta t H_1^{(k)}} \ket{\psi(0)}$
      \Statex \emph{Compute the QFI Eq.~\eqref{eq:QFI}}
      \State $F_r = F(\psi_r(T))$
      
      \EndFor
      \Statex \emph{Return the optimal trajectory}
      \State $r_{\max} = \arg\max_r \{F_r\}, \quad F_{\text{opt}} = F_{r_{\max}}$
          \Statex \begin{center}
          \textbf{Output}
      \end{center}
      \Statex \emph{Final Control Solution}
\State Optimal QFI $F_{\text{opt}}$ and corresponding control parameters $J_{i}^{(k)}, \Delta_{i}^{(k)}$ and $U_{i}^{(k)}$
\end{algorithmic}
\end{algorithm}

In Fig.~\ref{Fig_monte_carlo}, we show results of a Monte-Carlo optimization with $\nu=10$. In Fig.~\ref{Fig_monte_carlo}a, we show the maximized QFI for $L=3, 5$ for $N$ up to $200$. The data shows an intermediate scaling (solid lines). This is confirmed further for larger $L$, up to $10$ albeit for smaller $N$ ($10$ and $20$) in Fig.~\ref{Fig_monte_carlo}b. In Fig.~\ref{Fig_monte_carlo}c, we show the corresponding precision $\Delta\phi$ (Eq.~\eqref{eq:d_phi}). The value of $T$ is critical to generate sufficient entanglement in the system so that the statistics Eq.~(\ref{mean_std}) are valid for $\ket{\psi(T)}$. This is illustrated in Fig.~\ref{Fig_monte_carlo}d.

The Monte-Carlo optimization results in different scalings for the two cases considered.
\begin{itemize}
    \item[i.] $L=\mathcal{O}(1)$ and $N\gg 1$. In this case $d=\mathcal{O}(N)$ and $\mu = \mathcal{O}(N^2), \sigma=\mathcal{O}(N^{3/2})$. Therefore, $F_0=\mu-\sigma= \mathcal{O}(N^2)$ and one can already obtain a \textit{Heisenberg scaling} using a Monte-Carlo optimization.  See Fig.~\ref{Fig1}c, left panel.
    \item[ii.] $N, L\gg 1$. The dimension $d  \gg N, L$ and therefore, $1-1/\sqrt{d}=\mathcal{O}(1)$. $\mu =\mathcal{O}(N^2L+NL^2)$ and $\sigma = \mathcal{O}(\mu/\sqrt{d})$. Thus, the intermediate scaling $\mu-\sigma = \mathcal{O}(N^2L+NL^2)$ is easily accessible to Monte-Carlo optimization. While higher than standard quantum limit ($NL^2$), the intermediate scaling is lower than Heisenberg limit ($N^2L^2$). See Fig.~\ref{Fig1}c, right panel. The densities also suggest that it is non-trivial to find a trajectory with Heisenberg scaling. For $F_0 =\mathcal{O}(N^2L^2)$, $F_0-\mu = \mathcal{O}(N^2L^2)\gg \sigma $ and therefore,  $$p(F_0) \leq  \frac{\sigma^2}{(F_0-\mu)^2} = O\left(\frac{1}{d}\left( \frac{1}{L}+\frac{1}{N}\right)\right)$$
    which is vanishingly small.
\end{itemize}

While for small $L$ the intermediate scaling, which can be achieved using the Monte-Carlo optimization, is similar to Heisenberg scaling, they are significantly different for larger $L$. In fact, when $L\gg 1$, the set of states with a Heisenberg scaling has a vanishingly small measure. Therefore, one cannot find such trajectories using a Monte-Carlo optimization. Thus, optimizing the QFI up to the Heisenberg scaling in this case requires more sophisticated QOC techniques.

The case $N\gg L=\mathcal{O}(1)$ is relevant to state of the art experiments. In the next section, we outline possible experimental techniques to implement the Monte-Carlo optimization for this regime. 

\section{Experimental considerations}\label{Expt}
We discuss possible experimental schemes to implement a lattice with $L$ sites and  $N\gg L=\mathcal{O}(1)$ bosons. We also discuss how the Hamiltonian, Eq.~(\ref{eq:H}) can be implemented along with the tunability of the parameters.

A standard optical lattice formed by standing waves produced by back reflecting a laser allows for tuning of the tunneling rate and interactions globally, i.e., $J=J_{i, i+1}$ can be tuned while $J_{i, i+1}$ cannot be tuned independently. And so is the case with the Hubbard interactions. Moreover, in a standard optical lattice the occupancy per site, $n_i$ is limited to $2$ or $3$~\cite{Covey2016}. 
\begin{figure}[t]
    \centering
    \includegraphics[width=0.95\linewidth]{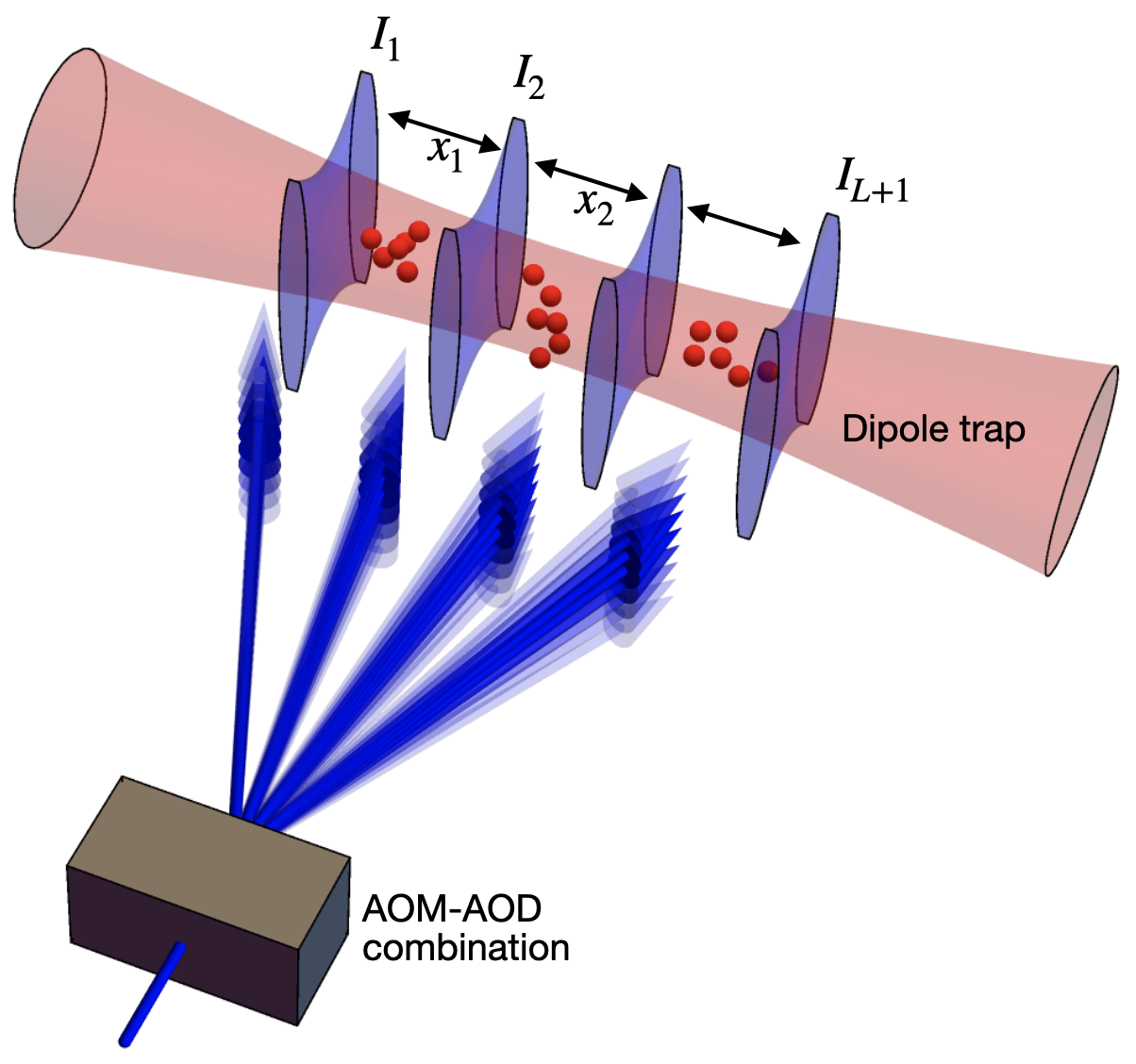}
    \caption{\textbf{Possible experimental design:} In contrast to a standard optical lattice formed out of standing waves, we propose a lattice formed out of repulsive barriers on an attractive optical dipole trap. The walls can be produced by painting a tweezer array, where the intensity of each wall and the separations can be tuned independently. }
    \label{Fig3}
\end{figure}

An alternate scheme to produce a lattice is to use a blue-detuned (i.e., repulsive) laser to ``paint" walls into a standard optical dipole trap (Fig.~\ref{Fig3}). The dipole trap can be produced using a red-detuned (i.e., attractive) laser focused to a waist of $\sim 20 \mu m$. We can then use a highly focused, diffraction limited blue-detuned laser to paint walls~\cite{henderson2009experimental}. We estimate the waist of the blue-detuned laser to be $\sim 0.5 \mu m$, which is also the thickness of the wall. One can use an acousto-optic deflector (AOD) to split the blue-detuned laser into a tweezer array i.e., an array of $L+1$ beams, where the intensity and position of each beam can be independently controlled using the RF input to the AOD. This allows us to independently control each of the intensities $I_j$ and the separations $x_j$. Finally, an acousto-optic modulator (AOM) can be used to sweep the beams vertically in order to paint a wall. 

The intensities $I_j$ control the tunneling rates in Eq.~(\ref{eq:H}), i.e., $J_{i, i+1}\sim \exp(-\sqrt{I_i})$ and the separations $x_i$ control the volume of each trapping site independently and therefore allow us to control the s-wave scattering interactions, $U_i\sim \frac{1}{x_i}$. Moreover, this lattice can have a larger volume per trapping site, compared to a standard optical lattice and therefore, allow larger per-site occupancies $n_i$. A second, attractive tweezer array can be used to independently tune the local potentials $\Delta_i$. 

One can engineer the system to obtain a typical hopping rate of $J\sim 100 \text{ Hz}$~\cite{ryu2020quantum, PhysRevLett.111.205301}. That sets the scale of $J_{i,i+1}$, $U_{i}$, $\Delta_i$ and $T$. Under the resulting scale, $T=20$, which is the maximum times used in the computations in Fig.~\ref{Fig3}, corresponds to $T=200\text{ ms}  (= 20\times 1/J)$, which is much shorter than the lifetime of BECs, which can be several $10$s of seconds~\cite{PhysRevLett.88.020401}. Both $\Delta$ and $U$ can be engineered to be a few $100 \text{Hz}$ using Feshbach resonance~\cite{RevModPhys.82.1225} and optical tweezers with independently controllable intensities~\cite{kaufman2021quantum}. The pulsed structure of the control in Eq.~\eqref{H1H2H3} can be engineered in state-of-the art experiments via pulse shaping~\cite{PhysRevApplied.13.054060}. 

The QFI of the Monte-Carlo optimized pulse sequences are expected to be robust to experimental fluctuations. The intermediate QFI scaling is the narrow peak of the distribution of QFI. Therefore, a vast majority of the perturbations of a state that has the intermediate scaling also have similar, intermediate scaling. Therefore, while the quantum state is not robust in general to experimental fluctuations, the QFI is expected to be relatively robust.

\section{Conclusion}\label{sec:conclusion}
We have demonstrated that the Monte-Carlo method provides a powerful and experimentally feasible framework for preparing multi-mode bosonic squeezed states, where intermediate scaling for large systems ($L = \mathcal{O}(1), N \gg L$) can be achieved. These results indicate that metrologically useful multi-mode entanglement can be generated using accessible experimental parameters, opening new directions for quantum-enhanced sensing in platforms such as Bose–Einstein condensates and optical lattices. High levels of control of spin-1 condensates, i.e., $L=3$ and $N\sim 10^4$ have been experimentally achieved~\cite{StamperKurn2013} and used to demonstrate various phenomena including  coherent spin-mixing oscillations among the three Zeeman sublevels ($m_F = +1, 0, -1$)~\cite{Chang2005}.

$L=5$ component BECs have also been studied experimentally using spin-2 systems~\cite{Tojo2009, Eto2018}. Combining internal spin states and spatial modes (i.e., lattices) has also been achieved~\cite{Shchedrin2018}. Optical lattices have been used to study coherent quantum many-body dynamics~\cite{Greiner2002, Morsch2006}. In all of these experimental works, the system is characterized by $L=\mathcal{O}(1)$ and $N\gg L$ --- the ideal regime to apply our results regarding the intermediate scaling. Therefore, our result can be implemented in these state-of-the-art experimental platforms.  

The technique presented here also applies to generalized and modified versions of the QFI. In particular, the mixed-state QFI is not smooth with respect to the quantum state~\cite{PhysRevA.95.052320}, making it difficult to optimize with gradient-based QOC methods. The Monte-Carlo approach introduced here provides a viable alternative. Optimizing multi-parameter QFI involves additional constraints related to compatibility~\cite{PhysRevA.94.052108} and is also a potential application of the Monte-Carlo technique. Finally, QFI and the Cramer-Rao bounds presents a lower bound on the precision and saturating it may require measuring in a non-trivial basis which may not be possible within the experimental constraints. To address this issue, one can construct a space of experimentally feasible observables, project the quantum state onto it and define the QFI within the projection~\cite{PRXQuantum.4.040305, PhysRevA.111.022436}. Understanding the statistical properties of such projected QFI can help us use Monte-Carlo optimization for this case as well.

While the ultimate scaling for small systems can be achieved via QOC (see sec.~\ref{sec:QOC} of the supplementary information~\cite{supplements}), the deterministic optimization method becomes inefficient once the system size increases. Future work includes implementation of both Monte-Carlo and QOC methods with tensor networks, extension the optimization techniques to higher-dimensional lattices, and experimental validation for precision gravity measurement.

\section*{Acknowledgements}
We thank Ivan Deutsch and Vikas Buchemmavari for fruitful discussions. Research presented in this article was supported by the Laboratory Directed Research and Development program of Los Alamos National Laboratory under project numbers 20230779PRD1, 20240396ER and 20260043DR.  Portions of this work were also supported by the U.S. Department of Energy, Office of Science, National Quantum Information Science Research Centers, Quantum Science Center (the computations related to Monte-Carlo optimization). 

\paragraph*{\textbf{Data and code availability}} The code used to produce the data on Monte-Carlo optimization and the data will be made available upon a reasonable request. 

\paragraph*{\textbf{Competing interests}} The authors declare no competing interests.

\bibliography{ref}

\begin{thebibliography}{57}%
\makeatletter
\providecommand \@ifxundefined [1]{%
 \@ifx{#1\undefined}
}%
\providecommand \@ifnum [1]{%
 \ifnum #1\expandafter \@firstoftwo
 \else \expandafter \@secondoftwo
 \fi
}%
\providecommand \@ifx [1]{%
 \ifx #1\expandafter \@firstoftwo
 \else \expandafter \@secondoftwo
 \fi
}%
\providecommand \natexlab [1]{#1}%
\providecommand \enquote  [1]{``#1''}%
\providecommand \bibnamefont  [1]{#1}%
\providecommand \bibfnamefont [1]{#1}%
\providecommand \citenamefont [1]{#1}%
\providecommand \href@noop [0]{\@secondoftwo}%
\providecommand \href [0]{\begingroup \@sanitize@url \@href}%
\providecommand \@href[1]{\@@startlink{#1}\@@href}%
\providecommand \@@href[1]{\endgroup#1\@@endlink}%
\providecommand \@sanitize@url [0]{\catcode `\\12\catcode `\$12\catcode `\&12\catcode `\#12\catcode `\^12\catcode `\_12\catcode `\%12\relax}%
\providecommand \@@startlink[1]{}%
\providecommand \@@endlink[0]{}%
\providecommand \url  [0]{\begingroup\@sanitize@url \@url }%
\providecommand \@url [1]{\endgroup\@href {#1}{\urlprefix }}%
\providecommand \urlprefix  [0]{URL }%
\providecommand \Eprint [0]{\href }%
\providecommand \doibase [0]{https://doi.org/}%
\providecommand \selectlanguage [0]{\@gobble}%
\providecommand \bibinfo  [0]{\@secondoftwo}%
\providecommand \bibfield  [0]{\@secondoftwo}%
\providecommand \translation [1]{[#1]}%
\providecommand \BibitemOpen [0]{}%
\providecommand \bibitemStop [0]{}%
\providecommand \bibitemNoStop [0]{.\EOS\space}%
\providecommand \EOS [0]{\spacefactor3000\relax}%
\providecommand \BibitemShut  [1]{\csname bibitem#1\endcsname}%
\let\auto@bib@innerbib\@empty
\bibitem [{\citenamefont {Montenegro}\ \emph {et~al.}(2025)\citenamefont {Montenegro}, \citenamefont {Mukhopadhyay}, \citenamefont {Yousefjani}, \citenamefont {Sarkar}, \citenamefont {Mishra}, \citenamefont {Paris},\ and\ \citenamefont {Bayat}}]{Montenegro_2025}%
  \BibitemOpen
  \bibfield  {author} {\bibinfo {author} {\bibfnamefont {V.}~\bibnamefont {Montenegro}}, \bibinfo {author} {\bibfnamefont {C.}~\bibnamefont {Mukhopadhyay}}, \bibinfo {author} {\bibfnamefont {R.}~\bibnamefont {Yousefjani}}, \bibinfo {author} {\bibfnamefont {S.}~\bibnamefont {Sarkar}}, \bibinfo {author} {\bibfnamefont {U.}~\bibnamefont {Mishra}}, \bibinfo {author} {\bibfnamefont {M.~G.}\ \bibnamefont {Paris}},\ and\ \bibinfo {author} {\bibfnamefont {A.}~\bibnamefont {Bayat}},\ }\bibfield  {title} {\bibinfo {title} {Review: Quantum metrology and sensing with many-body systems},\ }\href {https://doi.org/10.1016/j.physrep.2025.05.005} {\bibfield  {journal} {\bibinfo  {journal} {Physics Reports}\ }\textbf {\bibinfo {volume} {1134}},\ \bibinfo {pages} {1–62} (\bibinfo {year} {2025})}\BibitemShut {NoStop}%
\bibitem [{\citenamefont {Huang}\ \emph {et~al.}(2014)\citenamefont {Huang}, \citenamefont {Wu}, \citenamefont {Zhong},\ and\ \citenamefont {Lee}}]{Huang_2014}%
  \BibitemOpen
  \bibfield  {author} {\bibinfo {author} {\bibfnamefont {J.}~\bibnamefont {Huang}}, \bibinfo {author} {\bibfnamefont {S.}~\bibnamefont {Wu}}, \bibinfo {author} {\bibfnamefont {H.}~\bibnamefont {Zhong}},\ and\ \bibinfo {author} {\bibfnamefont {C.}~\bibnamefont {Lee}},\ }\bibinfo {title} {Quantum metrology with cold atoms},\ in\ \href {https://doi.org/10.1142/9789814590174_0007} {\emph {\bibinfo {booktitle} {Annual Review of Cold Atoms and Molecules}}}\ (\bibinfo  {publisher} {WORLD SCIENTIFIC},\ \bibinfo {year} {2014})\ p.\ \bibinfo {pages} {365–415}\BibitemShut {NoStop}%
\bibitem [{\citenamefont {Vengalattore}\ \emph {et~al.}(2007)\citenamefont {Vengalattore}, \citenamefont {Higbie}, \citenamefont {Leslie}, \citenamefont {Guzman}, \citenamefont {Sadler},\ and\ \citenamefont {Stamper-Kurn}}]{PhysRevLett.98.200801}%
  \BibitemOpen
  \bibfield  {author} {\bibinfo {author} {\bibfnamefont {M.}~\bibnamefont {Vengalattore}}, \bibinfo {author} {\bibfnamefont {J.~M.}\ \bibnamefont {Higbie}}, \bibinfo {author} {\bibfnamefont {S.~R.}\ \bibnamefont {Leslie}}, \bibinfo {author} {\bibfnamefont {J.}~\bibnamefont {Guzman}}, \bibinfo {author} {\bibfnamefont {L.~E.}\ \bibnamefont {Sadler}},\ and\ \bibinfo {author} {\bibfnamefont {D.~M.}\ \bibnamefont {Stamper-Kurn}},\ }\bibfield  {title} {\bibinfo {title} {High-resolution magnetometry with a spinor bose-einstein condensate},\ }\href {https://doi.org/10.1103/PhysRevLett.98.200801} {\bibfield  {journal} {\bibinfo  {journal} {Phys. Rev. Lett.}\ }\textbf {\bibinfo {volume} {98}},\ \bibinfo {pages} {200801} (\bibinfo {year} {2007})}\BibitemShut {NoStop}%
\bibitem [{\citenamefont {Gawlik}\ and\ \citenamefont {Higbie}(2013)}]{GawlikHigbie2013}%
  \BibitemOpen
  \bibfield  {author} {\bibinfo {author} {\bibfnamefont {W.}~\bibnamefont {Gawlik}}\ and\ \bibinfo {author} {\bibfnamefont {J.~M.}\ \bibnamefont {Higbie}},\ }\bibfield  {title} {\bibinfo {title} {Magnetometry with cold atoms},\ }in\ \href {https://doi.org/10.1017/CBO9780511846380.010} {\emph {\bibinfo {booktitle} {Optical Magnetometry}}},\ \bibinfo {editor} {edited by\ \bibinfo {editor} {\bibfnamefont {D.}~\bibnamefont {Budker}}\ and\ \bibinfo {editor} {\bibfnamefont {D.~F.~J.}\ \bibnamefont {Kimball}}}\ (\bibinfo  {publisher} {Cambridge University Press},\ \bibinfo {address} {Cambridge, UK},\ \bibinfo {year} {2013})\ pp.\ \bibinfo {pages} {167--189}\BibitemShut {NoStop}%
\bibitem [{\citenamefont {Kasevich}\ and\ \citenamefont {Chu}(1991)}]{PhysRevLett.67.181}%
  \BibitemOpen
  \bibfield  {author} {\bibinfo {author} {\bibfnamefont {M.}~\bibnamefont {Kasevich}}\ and\ \bibinfo {author} {\bibfnamefont {S.}~\bibnamefont {Chu}},\ }\bibfield  {title} {\bibinfo {title} {Atomic interferometry using stimulated raman transitions},\ }\href {https://doi.org/10.1103/PhysRevLett.67.181} {\bibfield  {journal} {\bibinfo  {journal} {Phys. Rev. Lett.}\ }\textbf {\bibinfo {volume} {67}},\ \bibinfo {pages} {181} (\bibinfo {year} {1991})}\BibitemShut {NoStop}%
\bibitem [{\citenamefont {Hald}\ \emph {et~al.}(1999)\citenamefont {Hald}, \citenamefont {S\o{}rensen}, \citenamefont {Schori},\ and\ \citenamefont {Polzik}}]{PhysRevLett.83.1319}%
  \BibitemOpen
  \bibfield  {author} {\bibinfo {author} {\bibfnamefont {J.}~\bibnamefont {Hald}}, \bibinfo {author} {\bibfnamefont {J.~L.}\ \bibnamefont {S\o{}rensen}}, \bibinfo {author} {\bibfnamefont {C.}~\bibnamefont {Schori}},\ and\ \bibinfo {author} {\bibfnamefont {E.~S.}\ \bibnamefont {Polzik}},\ }\bibfield  {title} {\bibinfo {title} {Spin squeezed atoms: A macroscopic entangled ensemble created by light},\ }\href {https://doi.org/10.1103/PhysRevLett.83.1319} {\bibfield  {journal} {\bibinfo  {journal} {Phys. Rev. Lett.}\ }\textbf {\bibinfo {volume} {83}},\ \bibinfo {pages} {1319} (\bibinfo {year} {1999})}\BibitemShut {NoStop}%
\bibitem [{\citenamefont {Est{\`e}ve}\ \emph {et~al.}(2008)\citenamefont {Est{\`e}ve}, \citenamefont {Gross}, \citenamefont {Weller}, \citenamefont {Giovanazzi},\ and\ \citenamefont {Oberthaler}}]{esteve2008squeezing}%
  \BibitemOpen
  \bibfield  {author} {\bibinfo {author} {\bibfnamefont {J.}~\bibnamefont {Est{\`e}ve}}, \bibinfo {author} {\bibfnamefont {C.}~\bibnamefont {Gross}}, \bibinfo {author} {\bibfnamefont {A.}~\bibnamefont {Weller}}, \bibinfo {author} {\bibfnamefont {S.}~\bibnamefont {Giovanazzi}},\ and\ \bibinfo {author} {\bibfnamefont {M.~K.}\ \bibnamefont {Oberthaler}},\ }\bibfield  {title} {\bibinfo {title} {Squeezing and entanglement in a bose--einstein condensate},\ }\href {https://doi.org/10.1038/nature07332} {\bibfield  {journal} {\bibinfo  {journal} {Nature}\ }\textbf {\bibinfo {volume} {455}},\ \bibinfo {pages} {1216} (\bibinfo {year} {2008})}\BibitemShut {NoStop}%
\bibitem [{\citenamefont {Hamley}\ \emph {et~al.}(2012{\natexlab{a}})\citenamefont {Hamley}, \citenamefont {Gerving}, \citenamefont {Hoang}, \citenamefont {Bookjans},\ and\ \citenamefont {Chapman}}]{hamley2012spin}%
  \BibitemOpen
  \bibfield  {author} {\bibinfo {author} {\bibfnamefont {C.~D.}\ \bibnamefont {Hamley}}, \bibinfo {author} {\bibfnamefont {C.~S.}\ \bibnamefont {Gerving}}, \bibinfo {author} {\bibfnamefont {T.~M.}\ \bibnamefont {Hoang}}, \bibinfo {author} {\bibfnamefont {E.~M.}\ \bibnamefont {Bookjans}},\ and\ \bibinfo {author} {\bibfnamefont {M.~S.}\ \bibnamefont {Chapman}},\ }\bibfield  {title} {\bibinfo {title} {Spin-nematic squeezed vacuum in a quantum gas},\ }\href {https://doi.org/10.1038/nphys2245} {\bibfield  {journal} {\bibinfo  {journal} {Nature Physics}\ }\textbf {\bibinfo {volume} {8}},\ \bibinfo {pages} {305} (\bibinfo {year} {2012}{\natexlab{a}})}\BibitemShut {NoStop}%
\bibitem [{\citenamefont {Leroux}\ \emph {et~al.}(2010)\citenamefont {Leroux}, \citenamefont {Schleier-Smith},\ and\ \citenamefont {Vuleti\ifmmode~\acute{c}\else \'{c}\fi{}}}]{PhysRevLett.104.073602}%
  \BibitemOpen
  \bibfield  {author} {\bibinfo {author} {\bibfnamefont {I.~D.}\ \bibnamefont {Leroux}}, \bibinfo {author} {\bibfnamefont {M.~H.}\ \bibnamefont {Schleier-Smith}},\ and\ \bibinfo {author} {\bibfnamefont {V.}~\bibnamefont {Vuleti\ifmmode~\acute{c}\else \'{c}\fi{}}},\ }\bibfield  {title} {\bibinfo {title} {Implementation of cavity squeezing of a collective atomic spin},\ }\href {https://doi.org/10.1103/PhysRevLett.104.073602} {\bibfield  {journal} {\bibinfo  {journal} {Phys. Rev. Lett.}\ }\textbf {\bibinfo {volume} {104}},\ \bibinfo {pages} {073602} (\bibinfo {year} {2010})}\BibitemShut {NoStop}%
\bibitem [{\citenamefont {Hosten}\ \emph {et~al.}(2016)\citenamefont {Hosten}, \citenamefont {Engelsen}, \citenamefont {Krishnakumar},\ and\ \citenamefont {Kasevich}}]{hosten2016measurement}%
  \BibitemOpen
  \bibfield  {author} {\bibinfo {author} {\bibfnamefont {O.}~\bibnamefont {Hosten}}, \bibinfo {author} {\bibfnamefont {N.~J.}\ \bibnamefont {Engelsen}}, \bibinfo {author} {\bibfnamefont {R.}~\bibnamefont {Krishnakumar}},\ and\ \bibinfo {author} {\bibfnamefont {M.~A.}\ \bibnamefont {Kasevich}},\ }\bibfield  {title} {\bibinfo {title} {Measurement noise 100 times lower than the quantum-projection limit using entangled atoms},\ }\href {https://doi.org/10.1038/nature16176} {\bibfield  {journal} {\bibinfo  {journal} {Nature}\ }\textbf {\bibinfo {volume} {529}},\ \bibinfo {pages} {505} (\bibinfo {year} {2016})}\BibitemShut {NoStop}%
\bibitem [{\citenamefont {Eckner}\ \emph {et~al.}(2023)\citenamefont {Eckner}, \citenamefont {Oppong}, \citenamefont {Cao}, \citenamefont {Young}, \citenamefont {Milner}, \citenamefont {Robinson}, \citenamefont {Ye},\ and\ \citenamefont {Kaufman}}]{eckner2023realizing}%
  \BibitemOpen
  \bibfield  {author} {\bibinfo {author} {\bibfnamefont {W.~J.}\ \bibnamefont {Eckner}}, \bibinfo {author} {\bibfnamefont {N.~D.}\ \bibnamefont {Oppong}}, \bibinfo {author} {\bibfnamefont {A.}~\bibnamefont {Cao}}, \bibinfo {author} {\bibfnamefont {A.~W.}\ \bibnamefont {Young}}, \bibinfo {author} {\bibfnamefont {W.~R.}\ \bibnamefont {Milner}}, \bibinfo {author} {\bibfnamefont {J.~M.}\ \bibnamefont {Robinson}}, \bibinfo {author} {\bibfnamefont {J.}~\bibnamefont {Ye}},\ and\ \bibinfo {author} {\bibfnamefont {A.~M.}\ \bibnamefont {Kaufman}},\ }\bibfield  {title} {\bibinfo {title} {Realizing spin squeezing with rydberg interactions in an optical clock},\ }\href {https://doi.org/10.1038/s41586-023-06434-6} {\bibfield  {journal} {\bibinfo  {journal} {Nature}\ }\textbf {\bibinfo {volume} {621}},\ \bibinfo {pages} {734} (\bibinfo {year} {2023})}\BibitemShut {NoStop}%
\bibitem [{\citenamefont {Sewell}\ \emph {et~al.}(2012)\citenamefont {Sewell}, \citenamefont {Koschorreck}, \citenamefont {Napolitano}, \citenamefont {Dubost}, \citenamefont {Behbood},\ and\ \citenamefont {Mitchell}}]{PhysRevLett.109.253605}%
  \BibitemOpen
  \bibfield  {author} {\bibinfo {author} {\bibfnamefont {R.~J.}\ \bibnamefont {Sewell}}, \bibinfo {author} {\bibfnamefont {M.}~\bibnamefont {Koschorreck}}, \bibinfo {author} {\bibfnamefont {M.}~\bibnamefont {Napolitano}}, \bibinfo {author} {\bibfnamefont {B.}~\bibnamefont {Dubost}}, \bibinfo {author} {\bibfnamefont {N.}~\bibnamefont {Behbood}},\ and\ \bibinfo {author} {\bibfnamefont {M.~W.}\ \bibnamefont {Mitchell}},\ }\bibfield  {title} {\bibinfo {title} {Magnetic sensitivity beyond the projection noise limit by spin squeezing},\ }\href {https://doi.org/10.1103/PhysRevLett.109.253605} {\bibfield  {journal} {\bibinfo  {journal} {Phys. Rev. Lett.}\ }\textbf {\bibinfo {volume} {109}},\ \bibinfo {pages} {253605} (\bibinfo {year} {2012})}\BibitemShut {NoStop}%
\bibitem [{\citenamefont {Gross}\ \emph {et~al.}(2010)\citenamefont {Gross}, \citenamefont {Zibold}, \citenamefont {Nicklas}, \citenamefont {Estève},\ and\ \citenamefont {Oberthaler}}]{gross2010nonlinear}%
  \BibitemOpen
  \bibfield  {author} {\bibinfo {author} {\bibfnamefont {C.}~\bibnamefont {Gross}}, \bibinfo {author} {\bibfnamefont {T.}~\bibnamefont {Zibold}}, \bibinfo {author} {\bibfnamefont {E.}~\bibnamefont {Nicklas}}, \bibinfo {author} {\bibfnamefont {J.}~\bibnamefont {Estève}},\ and\ \bibinfo {author} {\bibfnamefont {M.~K.}\ \bibnamefont {Oberthaler}},\ }\bibfield  {title} {\bibinfo {title} {Nonlinear atom interferometer surpasses classical precision limit},\ }\href {https://doi.org/10.1038/nature08919} {\bibfield  {journal} {\bibinfo  {journal} {Nature}\ }\textbf {\bibinfo {volume} {464}},\ \bibinfo {pages} {1165} (\bibinfo {year} {2010})}\BibitemShut {NoStop}%
\bibitem [{\citenamefont {Kitagawa}\ and\ \citenamefont {Ueda}(1993)}]{PhysRevA.47.5138}%
  \BibitemOpen
  \bibfield  {author} {\bibinfo {author} {\bibfnamefont {M.}~\bibnamefont {Kitagawa}}\ and\ \bibinfo {author} {\bibfnamefont {M.}~\bibnamefont {Ueda}},\ }\bibfield  {title} {\bibinfo {title} {Squeezed spin states},\ }\href {https://doi.org/10.1103/PhysRevA.47.5138} {\bibfield  {journal} {\bibinfo  {journal} {Phys. Rev. A}\ }\textbf {\bibinfo {volume} {47}},\ \bibinfo {pages} {5138} (\bibinfo {year} {1993})}\BibitemShut {NoStop}%
\bibitem [{\citenamefont {Vitagliano}\ \emph {et~al.}(2014)\citenamefont {Vitagliano}, \citenamefont {Apellaniz}, \citenamefont {Egusquiza},\ and\ \citenamefont {T\'oth}}]{PhysRevA.89.032307}%
  \BibitemOpen
  \bibfield  {author} {\bibinfo {author} {\bibfnamefont {G.}~\bibnamefont {Vitagliano}}, \bibinfo {author} {\bibfnamefont {I.}~\bibnamefont {Apellaniz}}, \bibinfo {author} {\bibfnamefont {I.~n.~L.}\ \bibnamefont {Egusquiza}},\ and\ \bibinfo {author} {\bibfnamefont {G.}~\bibnamefont {T\'oth}},\ }\bibfield  {title} {\bibinfo {title} {Spin squeezing and entanglement for an arbitrary spin},\ }\href {https://doi.org/10.1103/PhysRevA.89.032307} {\bibfield  {journal} {\bibinfo  {journal} {Phys. Rev. A}\ }\textbf {\bibinfo {volume} {89}},\ \bibinfo {pages} {032307} (\bibinfo {year} {2014})}\BibitemShut {NoStop}%
\bibitem [{\citenamefont {Tan}\ \emph {et~al.}(2020)\citenamefont {Tan}, \citenamefont {Huang}, \citenamefont {Xie},\ and\ \citenamefont {Wang}}]{PhysRevA.102.043307}%
  \BibitemOpen
  \bibfield  {author} {\bibinfo {author} {\bibfnamefont {Q.-S.}\ \bibnamefont {Tan}}, \bibinfo {author} {\bibfnamefont {Y.}~\bibnamefont {Huang}}, \bibinfo {author} {\bibfnamefont {Q.-T.}\ \bibnamefont {Xie}},\ and\ \bibinfo {author} {\bibfnamefont {X.}~\bibnamefont {Wang}},\ }\bibfield  {title} {\bibinfo {title} {Critically enhanced spin-nematic squeezing and entanglement in dipolar spinor condensates},\ }\href {https://doi.org/10.1103/PhysRevA.102.043307} {\bibfield  {journal} {\bibinfo  {journal} {Phys. Rev. A}\ }\textbf {\bibinfo {volume} {102}},\ \bibinfo {pages} {043307} (\bibinfo {year} {2020})}\BibitemShut {NoStop}%
\bibitem [{\citenamefont {Hamley}\ \emph {et~al.}(2012{\natexlab{b}})\citenamefont {Hamley}, \citenamefont {Gerving}, \citenamefont {Hoang}, \citenamefont {Bookjans},\ and\ \citenamefont {Chapman}}]{Hamley_2012}%
  \BibitemOpen
  \bibfield  {author} {\bibinfo {author} {\bibfnamefont {C.~D.}\ \bibnamefont {Hamley}}, \bibinfo {author} {\bibfnamefont {C.~S.}\ \bibnamefont {Gerving}}, \bibinfo {author} {\bibfnamefont {T.~M.}\ \bibnamefont {Hoang}}, \bibinfo {author} {\bibfnamefont {E.~M.}\ \bibnamefont {Bookjans}},\ and\ \bibinfo {author} {\bibfnamefont {M.~S.}\ \bibnamefont {Chapman}},\ }\bibfield  {title} {\bibinfo {title} {Spin-nematic squeezed vacuum in a quantum gas},\ }\href {https://doi.org/10.1038/nphys2245} {\bibfield  {journal} {\bibinfo  {journal} {Nature Physics}\ }\textbf {\bibinfo {volume} {8}},\ \bibinfo {pages} {305–308} (\bibinfo {year} {2012}{\natexlab{b}})}\BibitemShut {NoStop}%
\bibitem [{\citenamefont {Huang}\ \emph {et~al.}(2017)\citenamefont {Huang}, \citenamefont {Sun}, \citenamefont {Zhang}, \citenamefont {He},\ and\ \citenamefont {Sun}}]{Huang_2017}%
  \BibitemOpen
  \bibfield  {author} {\bibinfo {author} {\bibfnamefont {X.~Y.}\ \bibnamefont {Huang}}, \bibinfo {author} {\bibfnamefont {F.~X.}\ \bibnamefont {Sun}}, \bibinfo {author} {\bibfnamefont {W.}~\bibnamefont {Zhang}}, \bibinfo {author} {\bibfnamefont {Q.~Y.}\ \bibnamefont {He}},\ and\ \bibinfo {author} {\bibfnamefont {C.~P.}\ \bibnamefont {Sun}},\ }\bibfield  {title} {\bibinfo {title} {Spin-orbit-coupling-induced spin squeezing in three-component bose gases},\ }\bibfield  {journal} {\bibinfo  {journal} {Physical Review A}\ }\textbf {\bibinfo {volume} {95}},\ \href {https://doi.org/10.1103/physreva.95.013605} {10.1103/physreva.95.013605} (\bibinfo {year} {2017})\BibitemShut {NoStop}%
\bibitem [{\citenamefont {Brif}\ \emph {et~al.}(2010)\citenamefont {Brif}, \citenamefont {Chakrabarti},\ and\ \citenamefont {Rabitz}}]{Brif2010}%
  \BibitemOpen
  \bibfield  {author} {\bibinfo {author} {\bibfnamefont {C.}~\bibnamefont {Brif}}, \bibinfo {author} {\bibfnamefont {R.}~\bibnamefont {Chakrabarti}},\ and\ \bibinfo {author} {\bibfnamefont {H.}~\bibnamefont {Rabitz}},\ }\bibfield  {title} {\bibinfo {title} {Control of quantum phenomena: past, present and future},\ }\href {https://doi.org/10.1088/1367-2630/12/7/075008} {\bibfield  {journal} {\bibinfo  {journal} {New J. of Phys.}\ }\textbf {\bibinfo {volume} {12}},\ \bibinfo {pages} {075008} (\bibinfo {year} {2010})}\BibitemShut {NoStop}%
\bibitem [{\citenamefont {Dong}\ and\ \citenamefont {Petersen}(2010)}]{Dong2010}%
  \BibitemOpen
  \bibfield  {author} {\bibinfo {author} {\bibfnamefont {D.}~\bibnamefont {Dong}}\ and\ \bibinfo {author} {\bibfnamefont {I.}~\bibnamefont {Petersen}},\ }\bibfield  {title} {\bibinfo {title} {Quantum control theory and applications: a survey},\ }\href {https://doi.org/10.1049/iet-cta.2009.0508} {\bibfield  {journal} {\bibinfo  {journal} {EIT Control Theory Appl.}\ }\textbf {\bibinfo {volume} {4}},\ \bibinfo {pages} {2651} (\bibinfo {year} {2010})}\BibitemShut {NoStop}%
\bibitem [{\citenamefont {Altafini}\ and\ \citenamefont {Ticozzi}(2012)}]{Altafini2012}%
  \BibitemOpen
  \bibfield  {author} {\bibinfo {author} {\bibfnamefont {C.}~\bibnamefont {Altafini}}\ and\ \bibinfo {author} {\bibfnamefont {F.}~\bibnamefont {Ticozzi}},\ }\bibfield  {title} {\bibinfo {title} {Modeling and control of quantum systems: An introduction},\ }\href {https://doi.org/10.1109/TAC.2012.2195830} {\bibfield  {journal} {\bibinfo  {journal} {IEEE Trans. Autom. Control}\ }\textbf {\bibinfo {volume} {57}},\ \bibinfo {pages} {1898} (\bibinfo {year} {2012})}\BibitemShut {NoStop}%
\bibitem [{\citenamefont {Glaser}\ \emph {et~al.}(2015)\citenamefont {Glaser}, \citenamefont {Boscain}, \citenamefont {Calarco}, \citenamefont {Koch}, \citenamefont {K\"ockenberger}, \citenamefont {Kosloff}, \citenamefont {Kuprov}, \citenamefont {Luy}, \citenamefont {Schirmer}, \citenamefont {Schulte-Herbr\"uggen}, \citenamefont {Sugny},\ and\ \citenamefont {Wilhelm}}]{Q_Opt2015}%
  \BibitemOpen
  \bibfield  {author} {\bibinfo {author} {\bibfnamefont {S.~J.}\ \bibnamefont {Glaser}}, \bibinfo {author} {\bibfnamefont {U.}~\bibnamefont {Boscain}}, \bibinfo {author} {\bibfnamefont {T.}~\bibnamefont {Calarco}}, \bibinfo {author} {\bibfnamefont {C.~P.}\ \bibnamefont {Koch}}, \bibinfo {author} {\bibfnamefont {W.}~\bibnamefont {K\"ockenberger}}, \bibinfo {author} {\bibfnamefont {R.}~\bibnamefont {Kosloff}}, \bibinfo {author} {\bibfnamefont {I.}~\bibnamefont {Kuprov}}, \bibinfo {author} {\bibfnamefont {B.}~\bibnamefont {Luy}}, \bibinfo {author} {\bibfnamefont {S.}~\bibnamefont {Schirmer}}, \bibinfo {author} {\bibfnamefont {T.}~\bibnamefont {Schulte-Herbr\"uggen}}, \bibinfo {author} {\bibfnamefont {D.}~\bibnamefont {Sugny}},\ and\ \bibinfo {author} {\bibfnamefont {F.~K.}\ \bibnamefont {Wilhelm}},\ }\bibfield  {title} {\bibinfo {title} {Training {S}chrödinger’s cat: quantum optimal control},\ }\href@noop {} {\bibfield  {journal} {\bibinfo  {journal} {European Physics Journal D}\ }\textbf {\bibinfo {volume}
  {69}},\ \bibinfo {pages} {279} (\bibinfo {year} {2015})}\BibitemShut {NoStop}%
\bibitem [{\citenamefont {Khaneja}\ \emph {et~al.}(2005)\citenamefont {Khaneja}, \citenamefont {Reiss}, \citenamefont {Kehlet}, \citenamefont {Schulte-Herbr\"uggen},\ and\ \citenamefont {Glaser}}]{GRAPE}%
  \BibitemOpen
  \bibfield  {author} {\bibinfo {author} {\bibfnamefont {N.}~\bibnamefont {Khaneja}}, \bibinfo {author} {\bibfnamefont {T.}~\bibnamefont {Reiss}}, \bibinfo {author} {\bibfnamefont {C.}~\bibnamefont {Kehlet}}, \bibinfo {author} {\bibfnamefont {T.}~\bibnamefont {Schulte-Herbr\"uggen}},\ and\ \bibinfo {author} {\bibfnamefont {S.~J.}\ \bibnamefont {Glaser}},\ }\bibfield  {title} {\bibinfo {title} {Optimal control of coupled spin dynamics: design of {NMR} pulse sequences by gradient ascent algorithms},\ }\href {https://doi.org/10.1016/j.jmr.2004.11.004} {\bibfield  {journal} {\bibinfo  {journal} {J. Magnet. Res.}\ }\textbf {\bibinfo {volume} {172}},\ \bibinfo {pages} {296} (\bibinfo {year} {2005})}\BibitemShut {NoStop}%
\bibitem [{\citenamefont {Khaneja}\ \emph {et~al.}(2001)\citenamefont {Khaneja}, \citenamefont {Brockett},\ and\ \citenamefont {Glaser}}]{Khaneja_Brockett_2001}%
  \BibitemOpen
  \bibfield  {author} {\bibinfo {author} {\bibfnamefont {N.}~\bibnamefont {Khaneja}}, \bibinfo {author} {\bibfnamefont {R.}~\bibnamefont {Brockett}},\ and\ \bibinfo {author} {\bibfnamefont {S.~J.}\ \bibnamefont {Glaser}},\ }\bibfield  {title} {\bibinfo {title} {Time optimal control in spin systems},\ }\href {https://doi.org/10.1103/PhysRevA.63.032308} {\bibfield  {journal} {\bibinfo  {journal} {Phys. Rev. A}\ }\textbf {\bibinfo {volume} {63}},\ \bibinfo {pages} {032308} (\bibinfo {year} {2001})}\BibitemShut {NoStop}%
\bibitem [{\citenamefont {Sklarz}\ and\ \citenamefont {Tannor}(2002)}]{Krotov1}%
  \BibitemOpen
  \bibfield  {author} {\bibinfo {author} {\bibfnamefont {S.~E.}\ \bibnamefont {Sklarz}}\ and\ \bibinfo {author} {\bibfnamefont {D.~J.}\ \bibnamefont {Tannor}},\ }\bibfield  {title} {\bibinfo {title} {Loading a {Bose-Einstein} condensate onto an optical lattice: an application of optimal control theory to the nonlinear {S}chr\"odinger equation},\ }\href {https://doi.org/10.1103/PhysRevA.66.053619} {\bibfield  {journal} {\bibinfo  {journal} {Phys. Rev. A}\ }\textbf {\bibinfo {volume} {66}},\ \bibinfo {pages} {053619} (\bibinfo {year} {2002})}\BibitemShut {NoStop}%
\bibitem [{\citenamefont {Reich}\ \emph {et~al.}(2012)\citenamefont {Reich}, \citenamefont {Ndong},\ and\ \citenamefont {Koch}}]{Krotov2}%
  \BibitemOpen
  \bibfield  {author} {\bibinfo {author} {\bibfnamefont {D.~M.}\ \bibnamefont {Reich}}, \bibinfo {author} {\bibfnamefont {M.}~\bibnamefont {Ndong}},\ and\ \bibinfo {author} {\bibfnamefont {C.~P.}\ \bibnamefont {Koch}},\ }\bibfield  {title} {\bibinfo {title} {Monotonically convergent optimization in quantum control using {K}rotov's method},\ }\href {https://doi.org/10.1063/1.3691827} {\bibfield  {journal} {\bibinfo  {journal} {J. Chem. Phys.}\ }\textbf {\bibinfo {volume} {136}},\ \bibinfo {pages} {104103} (\bibinfo {year} {2012})}\BibitemShut {NoStop}%
\bibitem [{\citenamefont {Shao}\ \emph {et~al.}(2022)\citenamefont {Shao}, \citenamefont {Combes}, \citenamefont {Hauser},\ and\ \citenamefont {Nicotra}}]{Shao2022}%
  \BibitemOpen
  \bibfield  {author} {\bibinfo {author} {\bibfnamefont {J.}~\bibnamefont {Shao}}, \bibinfo {author} {\bibfnamefont {J.}~\bibnamefont {Combes}}, \bibinfo {author} {\bibfnamefont {J.}~\bibnamefont {Hauser}},\ and\ \bibinfo {author} {\bibfnamefont {M.~M.}\ \bibnamefont {Nicotra}},\ }\bibfield  {title} {\bibinfo {title} {Projection-operator-based newton method for the trajectory optimization of closed quantum systems},\ }\href {https://doi.org/10.1103/PhysRevA.105.032605} {\bibfield  {journal} {\bibinfo  {journal} {Phys. Rev. A}\ }\textbf {\bibinfo {volume} {105}},\ \bibinfo {pages} {032605} (\bibinfo {year} {2022})}\BibitemShut {NoStop}%
\bibitem [{\citenamefont {Shao}\ \emph {et~al.}(2024)\citenamefont {Shao}, \citenamefont {Naris}, \citenamefont {Hauser},\ and\ \citenamefont {Nicotra}}]{Shao2024}%
  \BibitemOpen
  \bibfield  {author} {\bibinfo {author} {\bibfnamefont {J.}~\bibnamefont {Shao}}, \bibinfo {author} {\bibfnamefont {M.}~\bibnamefont {Naris}}, \bibinfo {author} {\bibfnamefont {J.}~\bibnamefont {Hauser}},\ and\ \bibinfo {author} {\bibfnamefont {M.~M.}\ \bibnamefont {Nicotra}},\ }\bibfield  {title} {\bibinfo {title} {Solving quantum optimal control problems using projection-operator-based newton steps},\ }\href {https://doi.org/10.1103/PhysRevA.109.012609} {\bibfield  {journal} {\bibinfo  {journal} {Phys. Rev. A}\ }\textbf {\bibinfo {volume} {109}},\ \bibinfo {pages} {012609} (\bibinfo {year} {2024})}\BibitemShut {NoStop}%
\bibitem [{\citenamefont {Carrera}\ \emph {et~al.}(2025)\citenamefont {Carrera}, \citenamefont {Erbin},\ and\ \citenamefont {Misguich}}]{carrera2025preparingspinsqueezedstatesrydberg}%
  \BibitemOpen
  \bibfield  {author} {\bibinfo {author} {\bibfnamefont {E.~S.}\ \bibnamefont {Carrera}}, \bibinfo {author} {\bibfnamefont {H.}~\bibnamefont {Erbin}},\ and\ \bibinfo {author} {\bibfnamefont {G.}~\bibnamefont {Misguich}},\ }\href {https://arxiv.org/abs/2507.07875} {\bibinfo {title} {Preparing spin-squeezed states in rydberg atom arrays via quantum optimal control}} (\bibinfo {year} {2025}),\ \Eprint {https://arxiv.org/abs/2507.07875} {arXiv:2507.07875 [quant-ph]} \BibitemShut {NoStop}%
\bibitem [{\citenamefont {Layden}\ \emph {et~al.}(2023)\citenamefont {Layden}, \citenamefont {Mazzola}, \citenamefont {Mishmash}, \citenamefont {Motta}, \citenamefont {Wocjan}, \citenamefont {Kim},\ and\ \citenamefont {Sheldon}}]{layden2023quantum}%
  \BibitemOpen
  \bibfield  {author} {\bibinfo {author} {\bibfnamefont {D.}~\bibnamefont {Layden}}, \bibinfo {author} {\bibfnamefont {G.}~\bibnamefont {Mazzola}}, \bibinfo {author} {\bibfnamefont {R.~V.}\ \bibnamefont {Mishmash}}, \bibinfo {author} {\bibfnamefont {M.}~\bibnamefont {Motta}}, \bibinfo {author} {\bibfnamefont {P.}~\bibnamefont {Wocjan}}, \bibinfo {author} {\bibfnamefont {J.-S.}\ \bibnamefont {Kim}},\ and\ \bibinfo {author} {\bibfnamefont {S.}~\bibnamefont {Sheldon}},\ }\bibfield  {title} {\bibinfo {title} {Quantum-enhanced markov chain monte carlo},\ }\href@noop {} {\bibfield  {journal} {\bibinfo  {journal} {Nature}\ }\textbf {\bibinfo {volume} {619}},\ \bibinfo {pages} {282} (\bibinfo {year} {2023})}\BibitemShut {NoStop}%
\bibitem [{\citenamefont {Christmann}\ \emph {et~al.}(2025)\citenamefont {Christmann}, \citenamefont {Ivashkov}, \citenamefont {Chiurco},\ and\ \citenamefont {Mazzola}}]{christmann2025}%
  \BibitemOpen
  \bibfield  {author} {\bibinfo {author} {\bibfnamefont {J.}~\bibnamefont {Christmann}}, \bibinfo {author} {\bibfnamefont {P.}~\bibnamefont {Ivashkov}}, \bibinfo {author} {\bibfnamefont {M.}~\bibnamefont {Chiurco}},\ and\ \bibinfo {author} {\bibfnamefont {G.}~\bibnamefont {Mazzola}},\ }\bibfield  {title} {\bibinfo {title} {From quantum-enhanced to quantum-inspired monte carlo},\ }\href {https://doi.org/10.1103/PhysRevA.111.042615} {\bibfield  {journal} {\bibinfo  {journal} {Phys. Rev. A}\ }\textbf {\bibinfo {volume} {111}},\ \bibinfo {pages} {042615} (\bibinfo {year} {2025})}\BibitemShut {NoStop}%
\bibitem [{\citenamefont {Hryniuk}\ and\ \citenamefont {Szyma{\'{n}}ska}(2024)}]{Hryniuk2024tensornetworkbased}%
  \BibitemOpen
  \bibfield  {author} {\bibinfo {author} {\bibfnamefont {D.~A.}\ \bibnamefont {Hryniuk}}\ and\ \bibinfo {author} {\bibfnamefont {M.~H.}\ \bibnamefont {Szyma{\'{n}}ska}},\ }\bibfield  {title} {\bibinfo {title} {Tensor-network-based variational {M}onte {C}arlo approach to the non-equilibrium steady state of open quantum systems},\ }\href {https://doi.org/10.22331/q-2024-09-17-1475} {\bibfield  {journal} {\bibinfo  {journal} {{Quantum}}\ }\textbf {\bibinfo {volume} {8}},\ \bibinfo {pages} {1475} (\bibinfo {year} {2024})}\BibitemShut {NoStop}%
\bibitem [{\citenamefont {Henderson}\ \emph {et~al.}(2009)\citenamefont {Henderson}, \citenamefont {Ryu}, \citenamefont {MacCormick},\ and\ \citenamefont {Boshier}}]{henderson2009experimental}%
  \BibitemOpen
  \bibfield  {author} {\bibinfo {author} {\bibfnamefont {K.}~\bibnamefont {Henderson}}, \bibinfo {author} {\bibfnamefont {C.}~\bibnamefont {Ryu}}, \bibinfo {author} {\bibfnamefont {C.}~\bibnamefont {MacCormick}},\ and\ \bibinfo {author} {\bibfnamefont {M.~G.}\ \bibnamefont {Boshier}},\ }\bibfield  {title} {\bibinfo {title} {Experimental demonstration of painting arbitrary and dynamic potentials for bose--einstein condensates},\ }\href {https://iopscience.iop.org/article/10.1088/1367-2630/11/4/043030/meta} {\bibfield  {journal} {\bibinfo  {journal} {New Journal of Physics}\ }\textbf {\bibinfo {volume} {11}},\ \bibinfo {pages} {043030} (\bibinfo {year} {2009})}\BibitemShut {NoStop}%
\bibitem [{sup()}]{supplements}%
  \BibitemOpen
  \href@noop {} {}\bibinfo {note} {See Supplemental Material which includes Refs.~\cite{madhusudhana2023benchmarking1, BoucheronLugosiMassart2013, Feller1970, GrimmettStirzaker2001, Klenke2014}, not cited in the main text, for details on: derivation of the moments of the QFI, derivation of the bound on cumulative probability using the one-sided Chebychev inequality, a description of the code used for Monte-Carlo optimization and QOC for small systems.}\BibitemShut {Stop}%
\bibitem [{\citenamefont {Covey}\ \emph {et~al.}(2016)\citenamefont {Covey}, \citenamefont {Moses}, \citenamefont {G{\"a}rttner}, \citenamefont {Safavi-Naini}, \citenamefont {Miecnikowski}, \citenamefont {Fu}, \citenamefont {Schachenmayer}, \citenamefont {Julienne}, \citenamefont {Rey}, \citenamefont {Jin},\ and\ \citenamefont {Ye}}]{Covey2016}%
  \BibitemOpen
  \bibfield  {author} {\bibinfo {author} {\bibfnamefont {J.~P.}\ \bibnamefont {Covey}}, \bibinfo {author} {\bibfnamefont {S.~A.}\ \bibnamefont {Moses}}, \bibinfo {author} {\bibfnamefont {M.}~\bibnamefont {G{\"a}rttner}}, \bibinfo {author} {\bibfnamefont {A.}~\bibnamefont {Safavi-Naini}}, \bibinfo {author} {\bibfnamefont {M.~T.}\ \bibnamefont {Miecnikowski}}, \bibinfo {author} {\bibfnamefont {Z.}~\bibnamefont {Fu}}, \bibinfo {author} {\bibfnamefont {J.}~\bibnamefont {Schachenmayer}}, \bibinfo {author} {\bibfnamefont {P.~S.}\ \bibnamefont {Julienne}}, \bibinfo {author} {\bibfnamefont {A.~M.}\ \bibnamefont {Rey}}, \bibinfo {author} {\bibfnamefont {D.~S.}\ \bibnamefont {Jin}},\ and\ \bibinfo {author} {\bibfnamefont {J.}~\bibnamefont {Ye}},\ }\bibfield  {title} {\bibinfo {title} {Doublon dynamics and polar molecule production in an optical lattice},\ }\href {https://doi.org/10.1038/ncomms11279} {\bibfield  {journal} {\bibinfo  {journal} {Nature Communications}\ }\textbf {\bibinfo {volume} {7}},\ \bibinfo {pages}
  {11279} (\bibinfo {year} {2016})},\ \bibinfo {note} {article number: 11279}\BibitemShut {NoStop}%
\bibitem [{\citenamefont {Ryu}\ \emph {et~al.}(2020)\citenamefont {Ryu}, \citenamefont {Samson},\ and\ \citenamefont {Boshier}}]{ryu2020quantum}%
  \BibitemOpen
  \bibfield  {author} {\bibinfo {author} {\bibfnamefont {C.}~\bibnamefont {Ryu}}, \bibinfo {author} {\bibfnamefont {E.~C.}\ \bibnamefont {Samson}},\ and\ \bibinfo {author} {\bibfnamefont {M.~G.}\ \bibnamefont {Boshier}},\ }\bibfield  {title} {\bibinfo {title} {Quantum interference of currents in an atomtronic squid},\ }\href {https://doi.org/10.1038/s41467-020-17185-6} {\bibfield  {journal} {\bibinfo  {journal} {Nature Communications}\ }\textbf {\bibinfo {volume} {11}},\ \bibinfo {pages} {3338} (\bibinfo {year} {2020})}\BibitemShut {NoStop}%
\bibitem [{\citenamefont {Ryu}\ \emph {et~al.}(2013)\citenamefont {Ryu}, \citenamefont {Blackburn}, \citenamefont {Blinova},\ and\ \citenamefont {Boshier}}]{PhysRevLett.111.205301}%
  \BibitemOpen
  \bibfield  {author} {\bibinfo {author} {\bibfnamefont {C.}~\bibnamefont {Ryu}}, \bibinfo {author} {\bibfnamefont {P.~W.}\ \bibnamefont {Blackburn}}, \bibinfo {author} {\bibfnamefont {A.~A.}\ \bibnamefont {Blinova}},\ and\ \bibinfo {author} {\bibfnamefont {M.~G.}\ \bibnamefont {Boshier}},\ }\bibfield  {title} {\bibinfo {title} {Experimental realization of josephson junctions for an atom squid},\ }\href {https://doi.org/10.1103/PhysRevLett.111.205301} {\bibfield  {journal} {\bibinfo  {journal} {Phys. Rev. Lett.}\ }\textbf {\bibinfo {volume} {111}},\ \bibinfo {pages} {205301} (\bibinfo {year} {2013})}\BibitemShut {NoStop}%
\bibitem [{\citenamefont {Gustavson}\ \emph {et~al.}(2001)\citenamefont {Gustavson}, \citenamefont {Chikkatur}, \citenamefont {Leanhardt}, \citenamefont {G\"orlitz}, \citenamefont {Gupta}, \citenamefont {Pritchard},\ and\ \citenamefont {Ketterle}}]{PhysRevLett.88.020401}%
  \BibitemOpen
  \bibfield  {author} {\bibinfo {author} {\bibfnamefont {T.~L.}\ \bibnamefont {Gustavson}}, \bibinfo {author} {\bibfnamefont {A.~P.}\ \bibnamefont {Chikkatur}}, \bibinfo {author} {\bibfnamefont {A.~E.}\ \bibnamefont {Leanhardt}}, \bibinfo {author} {\bibfnamefont {A.}~\bibnamefont {G\"orlitz}}, \bibinfo {author} {\bibfnamefont {S.}~\bibnamefont {Gupta}}, \bibinfo {author} {\bibfnamefont {D.~E.}\ \bibnamefont {Pritchard}},\ and\ \bibinfo {author} {\bibfnamefont {W.}~\bibnamefont {Ketterle}},\ }\bibfield  {title} {\bibinfo {title} {Transport of bose-einstein condensates with optical tweezers},\ }\href {https://doi.org/10.1103/PhysRevLett.88.020401} {\bibfield  {journal} {\bibinfo  {journal} {Phys. Rev. Lett.}\ }\textbf {\bibinfo {volume} {88}},\ \bibinfo {pages} {020401} (\bibinfo {year} {2001})}\BibitemShut {NoStop}%
\bibitem [{\citenamefont {Chin}\ \emph {et~al.}(2010)\citenamefont {Chin}, \citenamefont {Grimm}, \citenamefont {Julienne},\ and\ \citenamefont {Tiesinga}}]{RevModPhys.82.1225}%
  \BibitemOpen
  \bibfield  {author} {\bibinfo {author} {\bibfnamefont {C.}~\bibnamefont {Chin}}, \bibinfo {author} {\bibfnamefont {R.}~\bibnamefont {Grimm}}, \bibinfo {author} {\bibfnamefont {P.}~\bibnamefont {Julienne}},\ and\ \bibinfo {author} {\bibfnamefont {E.}~\bibnamefont {Tiesinga}},\ }\bibfield  {title} {\bibinfo {title} {Feshbach resonances in ultracold gases},\ }\href {https://doi.org/10.1103/RevModPhys.82.1225} {\bibfield  {journal} {\bibinfo  {journal} {Rev. Mod. Phys.}\ }\textbf {\bibinfo {volume} {82}},\ \bibinfo {pages} {1225} (\bibinfo {year} {2010})}\BibitemShut {NoStop}%
\bibitem [{\citenamefont {Kaufman}\ and\ \citenamefont {Ni}(2021)}]{kaufman2021quantum}%
  \BibitemOpen
  \bibfield  {author} {\bibinfo {author} {\bibfnamefont {A.~M.}\ \bibnamefont {Kaufman}}\ and\ \bibinfo {author} {\bibfnamefont {K.-K.}\ \bibnamefont {Ni}},\ }\bibfield  {title} {\bibinfo {title} {Quantum science with optical tweezer arrays of ultracold atoms and molecules},\ }\href {https://doi.org/10.1038/s41567-021-01357-2} {\bibfield  {journal} {\bibinfo  {journal} {Nature Physics}\ }\textbf {\bibinfo {volume} {17}},\ \bibinfo {pages} {1324} (\bibinfo {year} {2021})}\BibitemShut {NoStop}%
\bibitem [{\citenamefont {Peterson}\ \emph {et~al.}(2020)\citenamefont {Peterson}, \citenamefont {Sarthour},\ and\ \citenamefont {Laflamme}}]{PhysRevApplied.13.054060}%
  \BibitemOpen
  \bibfield  {author} {\bibinfo {author} {\bibfnamefont {J.~P.}\ \bibnamefont {Peterson}}, \bibinfo {author} {\bibfnamefont {R.~S.}\ \bibnamefont {Sarthour}},\ and\ \bibinfo {author} {\bibfnamefont {R.}~\bibnamefont {Laflamme}},\ }\bibfield  {title} {\bibinfo {title} {Enhancing quantum control by improving shaped-pulse generation},\ }\href {https://doi.org/10.1103/PhysRevApplied.13.054060} {\bibfield  {journal} {\bibinfo  {journal} {Phys. Rev. Appl.}\ }\textbf {\bibinfo {volume} {13}},\ \bibinfo {pages} {054060} (\bibinfo {year} {2020})}\BibitemShut {NoStop}%
\bibitem [{\citenamefont {Stamper-Kurn}\ and\ \citenamefont {Ueda}(2013)}]{StamperKurn2013}%
  \BibitemOpen
  \bibfield  {author} {\bibinfo {author} {\bibfnamefont {D.~M.}\ \bibnamefont {Stamper-Kurn}}\ and\ \bibinfo {author} {\bibfnamefont {M.}~\bibnamefont {Ueda}},\ }\bibfield  {title} {\bibinfo {title} {Spinor bose gases: Symmetries, magnetism, and quantum dynamics},\ }\href {https://doi.org/10.1103/RevModPhys.85.1191} {\bibfield  {journal} {\bibinfo  {journal} {Rev. Mod. Phys.}\ }\textbf {\bibinfo {volume} {85}},\ \bibinfo {pages} {1191} (\bibinfo {year} {2013})}\BibitemShut {NoStop}%
\bibitem [{\citenamefont {Chang}\ \emph {et~al.}(2005)\citenamefont {Chang}, \citenamefont {Qin}, \citenamefont {Zhang}, \citenamefont {You},\ and\ \citenamefont {Chapman}}]{Chang2005}%
  \BibitemOpen
  \bibfield  {author} {\bibinfo {author} {\bibfnamefont {M.-S.}\ \bibnamefont {Chang}}, \bibinfo {author} {\bibfnamefont {Q.}~\bibnamefont {Qin}}, \bibinfo {author} {\bibfnamefont {W.}~\bibnamefont {Zhang}}, \bibinfo {author} {\bibfnamefont {L.}~\bibnamefont {You}},\ and\ \bibinfo {author} {\bibfnamefont {M.~S.}\ \bibnamefont {Chapman}},\ }\bibfield  {title} {\bibinfo {title} {Coherent spinor dynamics in a spin-1 bose condensate},\ }\href {https://doi.org/10.1038/nphys153} {\bibfield  {journal} {\bibinfo  {journal} {Nature Physics}\ }\textbf {\bibinfo {volume} {1}},\ \bibinfo {pages} {111–116} (\bibinfo {year} {2005})}\BibitemShut {NoStop}%
\bibitem [{\citenamefont {Tojo}\ \emph {et~al.}(2009)\citenamefont {Tojo}, \citenamefont {Hayashi}, \citenamefont {Tanabe}, \citenamefont {Hirano}, \citenamefont {Kawaguchi}, \citenamefont {Saito},\ and\ \citenamefont {Ueda}}]{Tojo2009}%
  \BibitemOpen
  \bibfield  {author} {\bibinfo {author} {\bibfnamefont {S.}~\bibnamefont {Tojo}}, \bibinfo {author} {\bibfnamefont {T.}~\bibnamefont {Hayashi}}, \bibinfo {author} {\bibfnamefont {T.}~\bibnamefont {Tanabe}}, \bibinfo {author} {\bibfnamefont {T.}~\bibnamefont {Hirano}}, \bibinfo {author} {\bibfnamefont {Y.}~\bibnamefont {Kawaguchi}}, \bibinfo {author} {\bibfnamefont {H.}~\bibnamefont {Saito}},\ and\ \bibinfo {author} {\bibfnamefont {M.}~\bibnamefont {Ueda}},\ }\bibfield  {title} {\bibinfo {title} {Spin-dependent inelastic collisions in spin-2 bose-einstein condensates},\ }\href {https://doi.org/10.1103/PhysRevA.80.042704} {\bibfield  {journal} {\bibinfo  {journal} {Phys. Rev. A}\ }\textbf {\bibinfo {volume} {80}},\ \bibinfo {pages} {042704} (\bibinfo {year} {2009})}\BibitemShut {NoStop}%
\bibitem [{\citenamefont {Eto}\ \emph {et~al.}(2018)\citenamefont {Eto}, \citenamefont {Shibayama}, \citenamefont {Saito},\ and\ \citenamefont {Hirano}}]{Eto2018}%
  \BibitemOpen
  \bibfield  {author} {\bibinfo {author} {\bibfnamefont {Y.}~\bibnamefont {Eto}}, \bibinfo {author} {\bibfnamefont {H.}~\bibnamefont {Shibayama}}, \bibinfo {author} {\bibfnamefont {H.}~\bibnamefont {Saito}},\ and\ \bibinfo {author} {\bibfnamefont {T.}~\bibnamefont {Hirano}},\ }\bibfield  {title} {\bibinfo {title} {Spinor dynamics in a mixture of spin-1 and spin-2 bose-einstein condensates},\ }\href {https://doi.org/10.1103/PhysRevA.97.021602} {\bibfield  {journal} {\bibinfo  {journal} {Phys. Rev. A}\ }\textbf {\bibinfo {volume} {97}},\ \bibinfo {pages} {021602} (\bibinfo {year} {2018})}\BibitemShut {NoStop}%
\bibitem [{\citenamefont {Shchedrin}\ \emph {et~al.}(2018)\citenamefont {Shchedrin}, \citenamefont {Jaschke},\ and\ \citenamefont {Carr}}]{Shchedrin2018}%
  \BibitemOpen
  \bibfield  {author} {\bibinfo {author} {\bibfnamefont {G.}~\bibnamefont {Shchedrin}}, \bibinfo {author} {\bibfnamefont {D.}~\bibnamefont {Jaschke}},\ and\ \bibinfo {author} {\bibfnamefont {L.~D.}\ \bibnamefont {Carr}},\ }\bibfield  {title} {\bibinfo {title} {Absence of landau damping in driven three-component bose–einstein condensate in optical lattices},\ }\bibfield  {journal} {\bibinfo  {journal} {Scientific Reports}\ }\textbf {\bibinfo {volume} {8}},\ \href {https://doi.org/10.1038/s41598-018-29454-y} {10.1038/s41598-018-29454-y} (\bibinfo {year} {2018})\BibitemShut {NoStop}%
\bibitem [{\citenamefont {Greiner}\ \emph {et~al.}(2002)\citenamefont {Greiner}, \citenamefont {Mandel}, \citenamefont {Esslinger}, \citenamefont {Hänsch},\ and\ \citenamefont {Bloch}}]{Greiner2002}%
  \BibitemOpen
  \bibfield  {author} {\bibinfo {author} {\bibfnamefont {M.}~\bibnamefont {Greiner}}, \bibinfo {author} {\bibfnamefont {O.}~\bibnamefont {Mandel}}, \bibinfo {author} {\bibfnamefont {T.}~\bibnamefont {Esslinger}}, \bibinfo {author} {\bibfnamefont {T.~W.}\ \bibnamefont {Hänsch}},\ and\ \bibinfo {author} {\bibfnamefont {I.}~\bibnamefont {Bloch}},\ }\bibfield  {title} {\bibinfo {title} {Quantum phase transition from a superfluid to a mott insulator in a gas of ultracold atoms},\ }\href {https://doi.org/10.1038/415039a} {\bibfield  {journal} {\bibinfo  {journal} {Nature}\ }\textbf {\bibinfo {volume} {415}},\ \bibinfo {pages} {39–44} (\bibinfo {year} {2002})}\BibitemShut {NoStop}%
\bibitem [{\citenamefont {Morsch}\ and\ \citenamefont {Oberthaler}(2006)}]{Morsch2006}%
  \BibitemOpen
  \bibfield  {author} {\bibinfo {author} {\bibfnamefont {O.}~\bibnamefont {Morsch}}\ and\ \bibinfo {author} {\bibfnamefont {M.}~\bibnamefont {Oberthaler}},\ }\bibfield  {title} {\bibinfo {title} {Dynamics of bose-einstein condensates in optical lattices},\ }\href {https://doi.org/10.1103/RevModPhys.78.179} {\bibfield  {journal} {\bibinfo  {journal} {Rev. Mod. Phys.}\ }\textbf {\bibinfo {volume} {78}},\ \bibinfo {pages} {179} (\bibinfo {year} {2006})}\BibitemShut {NoStop}%
\bibitem [{\citenamefont {\ifmmode~\check{S}\else \v{S}\fi{}afr\'anek}(2017)}]{PhysRevA.95.052320}%
  \BibitemOpen
  \bibfield  {author} {\bibinfo {author} {\bibfnamefont {D.}~\bibnamefont {\ifmmode~\check{S}\else \v{S}\fi{}afr\'anek}},\ }\bibfield  {title} {\bibinfo {title} {Discontinuities of the quantum fisher information and the bures metric},\ }\href {https://doi.org/10.1103/PhysRevA.95.052320} {\bibfield  {journal} {\bibinfo  {journal} {Phys. Rev. A}\ }\textbf {\bibinfo {volume} {95}},\ \bibinfo {pages} {052320} (\bibinfo {year} {2017})}\BibitemShut {NoStop}%
\bibitem [{\citenamefont {Ragy}\ \emph {et~al.}(2016)\citenamefont {Ragy}, \citenamefont {Jarzyna},\ and\ \citenamefont {Demkowicz-Dobrza\ifmmode~\acute{n}\else \'{n}\fi{}ski}}]{PhysRevA.94.052108}%
  \BibitemOpen
  \bibfield  {author} {\bibinfo {author} {\bibfnamefont {S.}~\bibnamefont {Ragy}}, \bibinfo {author} {\bibfnamefont {M.}~\bibnamefont {Jarzyna}},\ and\ \bibinfo {author} {\bibfnamefont {R.}~\bibnamefont {Demkowicz-Dobrza\ifmmode~\acute{n}\else \'{n}\fi{}ski}},\ }\bibfield  {title} {\bibinfo {title} {Compatibility in multiparameter quantum metrology},\ }\href {https://doi.org/10.1103/PhysRevA.94.052108} {\bibfield  {journal} {\bibinfo  {journal} {Phys. Rev. A}\ }\textbf {\bibinfo {volume} {94}},\ \bibinfo {pages} {052108} (\bibinfo {year} {2016})}\BibitemShut {NoStop}%
\bibitem [{\citenamefont {Zhou}\ \emph {et~al.}(2023)\citenamefont {Zhou}, \citenamefont {Michalakis},\ and\ \citenamefont {Gefen}}]{PRXQuantum.4.040305}%
  \BibitemOpen
  \bibfield  {author} {\bibinfo {author} {\bibfnamefont {S.}~\bibnamefont {Zhou}}, \bibinfo {author} {\bibfnamefont {S.}~\bibnamefont {Michalakis}},\ and\ \bibinfo {author} {\bibfnamefont {T.}~\bibnamefont {Gefen}},\ }\bibfield  {title} {\bibinfo {title} {Optimal protocols for quantum metrology with noisy measurements},\ }\href {https://doi.org/10.1103/PRXQuantum.4.040305} {\bibfield  {journal} {\bibinfo  {journal} {PRX Quantum}\ }\textbf {\bibinfo {volume} {4}},\ \bibinfo {pages} {040305} (\bibinfo {year} {2023})}\BibitemShut {NoStop}%
\bibitem [{\citenamefont {Liu}\ \emph {et~al.}(2025)\citenamefont {Liu}, \citenamefont {Yang}, \citenamefont {Shi},\ and\ \citenamefont {Yu}}]{PhysRevA.111.022436}%
  \BibitemOpen
  \bibfield  {author} {\bibinfo {author} {\bibfnamefont {J.-X.}\ \bibnamefont {Liu}}, \bibinfo {author} {\bibfnamefont {J.}~\bibnamefont {Yang}}, \bibinfo {author} {\bibfnamefont {H.-L.}\ \bibnamefont {Shi}},\ and\ \bibinfo {author} {\bibfnamefont {S.}~\bibnamefont {Yu}},\ }\bibfield  {title} {\bibinfo {title} {Optimal local measurements in single-parameter quantum metrology},\ }\href {https://doi.org/10.1103/PhysRevA.111.022436} {\bibfield  {journal} {\bibinfo  {journal} {Phys. Rev. A}\ }\textbf {\bibinfo {volume} {111}},\ \bibinfo {pages} {022436} (\bibinfo {year} {2025})}\BibitemShut {NoStop}%
\bibitem [{\citenamefont {Madhusudhana}(2023)}]{madhusudhana2023benchmarking1}%
  \BibitemOpen
  \bibfield  {author} {\bibinfo {author} {\bibfnamefont {B.~H.}\ \bibnamefont {Madhusudhana}},\ }\href@noop {} {\bibinfo {title} {Benchmarking multi-qubit gates-{I}: Metrological aspects}} (\bibinfo {year} {2023}),\ \Eprint {https://arxiv.org/abs/2210.04330} {arXiv:2210.04330} \BibitemShut {NoStop}%
\bibitem [{\citenamefont {Boucheron}\ \emph {et~al.}(2013)\citenamefont {Boucheron}, \citenamefont {Lugosi},\ and\ \citenamefont {Massart}}]{BoucheronLugosiMassart2013}%
  \BibitemOpen
  \bibfield  {author} {\bibinfo {author} {\bibfnamefont {S.}~\bibnamefont {Boucheron}}, \bibinfo {author} {\bibfnamefont {G.}~\bibnamefont {Lugosi}},\ and\ \bibinfo {author} {\bibfnamefont {P.}~\bibnamefont {Massart}},\ }\href@noop {} {\emph {\bibinfo {title} {Concentration Inequalities: A Nonasymptotic Theory of Independence}}}\ (\bibinfo  {publisher} {Oxford University Press},\ \bibinfo {address} {Oxford},\ \bibinfo {year} {2013})\BibitemShut {NoStop}%
\bibitem [{\citenamefont {Feller}(1968)}]{Feller1970}%
  \BibitemOpen
  \bibfield  {author} {\bibinfo {author} {\bibfnamefont {W.}~\bibnamefont {Feller}},\ }\href@noop {} {\emph {\bibinfo {title} {An Introduction to Probability Theory and Its Applications, Vol.~I}}},\ \bibinfo {edition} {3rd}\ ed.\ (\bibinfo  {publisher} {John Wiley \& Sons},\ \bibinfo {address} {New York},\ \bibinfo {year} {1968})\BibitemShut {NoStop}%
\bibitem [{\citenamefont {Grimmett}\ and\ \citenamefont {Stirzaker}(2001)}]{GrimmettStirzaker2001}%
  \BibitemOpen
  \bibfield  {author} {\bibinfo {author} {\bibfnamefont {G.}~\bibnamefont {Grimmett}}\ and\ \bibinfo {author} {\bibfnamefont {D.}~\bibnamefont {Stirzaker}},\ }\href@noop {} {\emph {\bibinfo {title} {Probability and Random Processes}}},\ \bibinfo {edition} {3rd}\ ed.\ (\bibinfo  {publisher} {Oxford University Press},\ \bibinfo {address} {Oxford},\ \bibinfo {year} {2001})\BibitemShut {NoStop}%
\bibitem [{\citenamefont {Klenke}(2014)}]{Klenke2014}%
  \BibitemOpen
  \bibfield  {author} {\bibinfo {author} {\bibfnamefont {A.}~\bibnamefont {Klenke}},\ }\href {https://doi.org/10.1007/978-1-4471-5361-0} {\emph {\bibinfo {title} {Probability Theory: A Comprehensive Course}}},\ \bibinfo {edition} {2nd}\ ed.\ (\bibinfo  {publisher} {Springer},\ \bibinfo {address} {London},\ \bibinfo {year} {2014})\BibitemShut {NoStop}%
\end{thebibliography}%

\cleardoublepage

\setcounter{figure}{0}
\setcounter{page}{1}
\setcounter{equation}{0}
\setcounter{section}{0}

\renewcommand{\thepage}{S\arabic{page}}
\renewcommand{\thesection}{S\arabic{section}}
\renewcommand{\theequation}{S\arabic{equation}}
\renewcommand{\thefigure}{S\arabic{figure}}
\onecolumngrid
\begin{center}
\huge{Supplementary Information}
\vspace{5mm}
\end{center}
\twocolumngrid
\normalsize

\section{Derivation of Eq.~(12) in the main text}\label{sec:mean_std}
The statistical properties of QFI, described by Eq.~(12) in the main text are crucial for many of the results presented here. We present the derivation of these two equations below.

Let us consider the number basis $\mathcal B = \{\ket{n_1,\cdots,n_L}: \sum_i n_i =N\}$ for $N$ bosons in $L$ modes. We represent the tuple $(n_1\cdots n_L)$ by $\vec{n}$. A state can be written in this basis as $\ket{\psi}=\sum_{\vec{n}}\psi_{\vec{n}}\ket{\vec{n}}$ where $\psi_{\vec{n}}$ are the coefficients. The operator $S_z=\sum_{\vec{n}}(\sum_i i\times n_i )\ket{\vec{n}}\bra{\vec{n}} $ is diagonal in this basis. We write $\sum_{i}i\times n_i = s_{\vec{n}}$. The QFI w.r.t $S_z$, $F(\psi)$ depends only on the probability weights, $p_{\vec{n}}=|\psi_{\vec{n}}|^2$:
\begin{equation}\label{QFI_apdx}
F(\psi) = 4\times \left(\sum_{\vec{n}}s^2_{\vec{n}} p_{\vec{n}}-\left(\sum_{\vec{n}}s_{\vec{n}} p_{\vec{n}}\right)^2\right)
\end{equation}

Computing the average of $F(\psi)$ and $F^2(\psi)$ over Haar random states requires moments $p_{\vec{n}}$ (e.g., $\int_{\text{Haar}} p_{\vec{n}}$, $\int_{\text{Haar}} p^2_{\vec{n}}$, $\int_{\text{Haar}} p_{\vec{n}}p_{\vec{m}}$ over Haar random states) and the sums of the form $\sum_{\vec{n}}s_{\vec{n}}^k$ for $k=1,2,3,4$. We will use the below tools.

\subsection{Moments of the Haar measure}
We represent the general moment $ \int_{\text{Haar}} p_{\vec{n}_1}^{r_1}\cdots p_{\vec{n}_{\ell}}^{r_{\ell}} $ for integers $r_1 ... r_{\ell}$ by $M_{r_1\cdots r_{\ell}}$ and use the results in ref.~\cite{madhusudhana2023benchmarking1} to compute it:
\begin{eqnarray}\label{genM}
    M_{r_1\cdots r_{\ell}}\!\!=\!\!\int_{\text{Haar}}\!\!\! p_{\vec{n}_1}^{r_1}\!\cdots p_{\vec{n}_{\ell}}^{r_{\ell}}\! = \!\frac{r_1!\cdots r_{\ell}! (d-1)!}{(d+ r_1 + \cdots + r_{\ell}-1)!}
\end{eqnarray}
where $d=\binom{N+L-1}{N}$ is the Hilbert space dimension. For instance, 
\begin{eqnarray}\label{Ms}
    \begin{split}
        M_1= \int_{\text{Haar}} p_{\vec{n}} &= \frac{1}{d}\\
        M_{2}=\int_{\text{Haar}} p_{\vec{n}}^2 &= \frac{2}{d(d+1)}\\
        M_{11}=\int_{\text{Haar}} p_{\vec{n}}p_{\vec{m}}&=\frac{1}{d(d+1)} \text{  if }(\vec{n}\neq \vec{m})
    \end{split}
\end{eqnarray}

\subsection{Sums of $s^k_{\vec{n}}$}
Next, we will compute the sums 
\begin{eqnarray}
    S_k = \sum_{\vec{n}} s^k_{\vec{n}}.
\end{eqnarray}
An important combinatorial generating function for bosonic systems is 
\begin{equation}
    f(t, x) = \frac{1}{1-tx} \frac{1}{1-tx^2}\cdots  \frac{1}{1-tx^L}
\end{equation}
For instance, the coefficient of $t^N$ in $f(t, 1)$ is the dimension $d(N, L)$ of the Hilbert space. Moreover, the coefficient of  $t^N$ in the derivative $\frac{\partial}{\partial x}f(t, x)$ at $x=1$ is equal to $\sum_{\sum n_i=N} s_{\vec{n}}$. In fact,
\begin{eqnarray}
    f(t, x) = \sum_{\vec{n}} t^{\sum n_i} x^{s(\vec{n})}
\end{eqnarray}
Therefore, all sums of the form $\sum_{\vec{n}}s_{\vec{n}}^k$ can be computed by taking derivatives of $f$. 

The first derivative can be computed via
\begin{equation}
    \begin{split}
        \partial_x \log f(t,x)
=&\sum_{i=1}^L \frac{t\,i\,x^{i-1}}{1-tx^i}\\
\partial_x f(t,x)=&f(t,x)\sum_{i=1}^L \frac{t\,i\,x^{i-1}}{1-tx^i}.
    \end{split}
\end{equation}

Evaluating at \(x=1\) gives
\begin{equation}
    \begin{split}
        \partial_x f(t,x)\big|_{x=1}
=&f(t,1)\sum_{i=1}^L \frac{t\,i}{1-t}\\
=&\frac{L(L+1)}{2}\frac{t}{(1-t)^{L+1}}.
    \end{split}
\end{equation}
Differentiating the RHS $N$ times, we obtain the coefficient
\begin{equation}
    \begin{split}
        \sum_{\sum n_i=N}s_{\vec{n}}
=&[t^N]\partial_x f(t,x)\big|_{x=1}\\
=&\frac{L(L+1)}{2}\,[t^{N-1}](1-t)^{-(L+1)}\\
=&\frac{L(L+1)}{2}\binom{L+N-1}{N-1}.
    \end{split}
\end{equation}
Since the dimension \(d=[t^N]f(t,1)=\binom{N+L-1}{L-1}\), this can also be written as
\begin{equation}\label{S1}
    S_1=\sum_{\sum n_i=N}s_{\vec{n}}
=\frac{L+1}{2}\,N\,d.
\end{equation}
For the second derivative, write
\[
A(t,x):=\sum_{i=1}^L \frac{t\,i\,x^{i-1}}{1-tx^i},\qquad
\partial_x^2 f=f\bigl(A^2+\partial_x A\bigr).
\]
Differentiate a single term of \(A\):
\[
\frac{d}{dx}\!\left(\frac{t\,i\,x^{i-1}}{1-tx^i}\right)
=\frac{t\,i\bigl((i-1)x^{i-2}(1-tx^i)+t\,i\,x^{2i-2}\bigr)}{(1-tx^i)^2},
\]
so at \(x=1\),
\begin{equation}
    \begin{split}
        \partial_x A(t,x)\big|_{x=1}
=&\frac{t}{(1-t)^2}\sum_{i=1}^L i\,(i-1+t)\\
=&\frac{t}{(1-t)^2}\!\left(\sum i(i-1)+t\sum i\right).
    \end{split}
\end{equation}

Using \(\sum_{i=1}^L i=\frac{L(L+1)}{2}\) and
\(\sum_{i=1}^L i(i-1)=\frac{L(L+1)(L-1)}{3}\), we obtain
\begin{equation}
    \begin{split}
        A(t,1)&=\frac{t\,L(L+1)}{2(1-t)},\\
\partial_x A(t,x)\big|_{x=1}
&=\frac{t\,L(L+1)}{(1-t)^2}\!\left(\frac{L-1}{3}+\frac{t}{2}\right).
    \end{split}
\end{equation}
Therefore
\begin{equation}
    \begin{split}
        &\partial_x^2 f(t,x)\big|_{x=1}=\\
&\frac{1}{(1-t)^L}\!\left(\frac{t\,L(L+1)}{2(1-t)}\right)^{\!2}\\
&+\frac{1}{(1-t)^L}
\frac{t\,L(L+1)}{(1-t)^2}\!\left(\frac{L-1}{3}+\frac{t}{2}\right).
    \end{split}
\end{equation}
A short algebraic simplification yields
\[
\partial_x^2 f(t,x)\big|_{x=1}
=\frac{L(L+1)}{(1-t)^{L+2}}
\left(\frac{2L+1}{6}\,t+\frac{L(L+1)}{4}\,t^2\right).
\]
Now using
\[
\sum_{\vec{n}}s_{\vec{n}}^2
=[t^N]\bigl(\partial_x^2 f+\partial_x f\bigr)\big|_{x=1}
\]

Extracting coefficients,
\begin{equation}\label{S2_1}
    \begin{split}
        S_2=&\sum_{\vec{n}}s_{\vec{n}}^2
= \\&L(L+1)\left[
\frac{2L+1}{6}\binom{L+N}{N-1}
+\frac{L(L+1)}{4}\binom{L+N-1}{N-2}
\right].
    \end{split}
\end{equation}
we can factor out the dimension \(d=\binom{N+L-1}{L-1}\) to obtain the compact form

\begin{eqnarray}\label{S2}
    S_2\!=\!\!\!\sum_{\vec{n}}\!s^2_{\vec{n}}\!
=\!d\!\left(\!\frac{3L^2+7L+2}{12}\!N^2\!+\!\frac{L(L-1)}{12}\!N\!\!\right).
\end{eqnarray}

Similarly, we can derive
\begin{equation}\label{S3}
    \begin{split}
        S_3=\sum_{\vec{n}}s^3_{\vec{n}} = d\left( \left(\frac{L^3}{8}+\frac{L^2}{2}+\frac{3L}{8}\right)N^3+\left(\frac{L^3}{8}-\frac{L}{8}\right)N^2\right)
    \end{split}
\end{equation}
And,
\begin{equation}\label{S4}
\begin{split}
   S_4= \sum_{\vec{n}}s_{\vec{n}}^4
=\frac{d}{240}\Big[
(15L^4+90L^3+125L^2+18L-8)N^4\\
+(30L^4+40L^3-54L^2-16L)\,N^3\\
+(5L^4-14L^3+11L^2-2L)\,N^2
-2L^2(L-1)^2\,N
\Big]
\end{split}
\end{equation}

\subsection{Mean of QFI}
Averaging Eq.~\eqref{QFI_apdx} over Haar random states, one obtains
\begin{eqnarray}
   \int_{\text{Haar}}\!\!\!\!\!\!F(\psi)\!\! =\! \mu\!\! =\! 4\!\times\! \left(\!M_1 S_2 \!-\! M_{11}\!(S_1^2\!-\!S_2)\! -\! M_2S_2 \!\right)
\end{eqnarray}
Here, $M$'s and $S$'s are the moments and sums defined above. Using $M_1=\frac{1}{d}$, $M_{11}=\frac{1}{d(d+1)}$ and $M_2=\frac{2}{d(d+1)}$ from Eq.~\eqref{Ms}, and Eq.~\eqref{S1} \& Eq.~\eqref{S2} for $S_1$ and $S_2$, and simplifying,
\begin{eqnarray}
    \mu 
= \frac{d}{d+1}\,\frac{N(L-1)(N+L)}{3}.
\end{eqnarray}

\subsection{Standard deviation of QFI}
We first compute the second moment, $\int_{\text{Haar}}F^2(\psi)$ of the QFI in terms of the sums and moments. 
\newcommand{\FQ}{F}
\newcommand{\avg}[1]{\left\langle #1 \right\rangle}

\newcommand{\Rcomb}{\Bigl(S_1^2-S_2\Bigr)S_2-2S_1S_3+2S_4}
\newcommand{\Dcomb}{S_1^4-6S_1^2S_2+3S_2^2+8S_1S_3-6S_4}

\begin{equation}
    \begin{split}
        \int_{\text{Haar}} F^2(\psi) &= 16\times \left[
S_4 M_2+(S_2^2-S_4)M_{11}\right. \\
&-2\left(S_4 M_3+(S_2^2+2S_1S_3-3S_4)M_{21}\right.+\\
&\left.\Rcomb\, M_{111}\right)\\
&+\;S_4 M_4+4(S_1S_3-S_4)M_{31}+3(S_2^2-S_4)M_{22}\\
&+6\,\Rcomb\, M_{211}\\
&+\Dcomb\, M_{1111}\left.\right]\\
    \end{split}
\end{equation}
Using Eq.~\eqref{genM}, we can simplify this to:

\begin{equation}
    \begin{split}
        \int_{\text{Haar}} F^2(\psi) &
=\frac{16}{d(d+1)(d+2)(d+3)}\times \\
&\Big(
S_1^4-2d\,S_1^2 S_2-(4d+4)\,S_1 S_3\\
&+(d^2+3d+3)\,S_2^2+(d^2+d)\,S_4
\Big).\
    \end{split}
\end{equation}
We then use Eq.~\eqref{S1}, Eq.~\eqref{S2}, Eq.~\eqref{S3}, Eq.~\eqref{S4} to substitute for the $S_k$s. The standard deviation can be obtained using the expression:
\begin{eqnarray}
    \sigma^2 = \int_{\text{Haar}} F^2(\psi) -\mu^2
\end{eqnarray}
The final expression we obtain is:
\begin{equation}
\begin{split}
        \sigma^2 
=& \frac{2\,N\,d\,(L-1)(N+L)}{45\,(d+1)^2(d+2)(d+3)}\\
&\times \Big\{\,N(N+L)\big[5L(d^2+3d+3)-(8d^2+21d+18)\big]\\
&-3L(L-1)(d+1)^2\Big\}.
\end{split}
\end{equation}
The leading order is presented in Eq.~(12) of the main text. 

\section{One–sided Chebyshev (Cantelli) inequality}\label{sec:chebyshev}
In this section, we provide a proof of the one–sided Chebyshev (Cantelli) inequality used in Eq.~(13) in the main text. We state the result before proving it. \\

\noindent \textbf{Theorem:} Let $X$ be a real random variable with mean $\mu$ and variance $\sigma^2<\infty$.
Then for all $t>0$,
\begin{equation}
    \begin{split}
        \Pr(X-\mu\ge t)\ &\le\ \frac{\sigma^2}{\sigma^2+t^2}
\qquad\text{and,}\\
\Pr(X-\mu\le -t)\ &\le\ \frac{\sigma^2}{\sigma^2+t^2}.
    \end{split}
\end{equation}

\vspace{0.3cm}

\noindent \textbf{Proof:} Let $Y:=X-\mu$ so $\langle Y\rangle =0$ and $\langle Y^2\rangle=\sigma^2$. Here, $\langle \cdot\rangle$ represents averaging over the probability distribution.
For any $\lambda\ge0$ and $t>0$,
\begin{equation}
    \begin{split}
        \Pr(Y\ge t)\ =&\ \Pr\!\big((Y+\lambda)^2\ge(t+\lambda)^2\big)
\ \\
\le\ &
\frac{\langle (Y+\lambda)^2\rangle }{(t+\lambda)^2}
\ =\ \frac{\sigma^2+\lambda^2}{(t+\lambda)^2},
    \end{split}
\end{equation}
by Markov’s inequality. Minimizing the right-hand side over $\lambda\ge0$ gives
$\lambda^\star=\sigma^2/t$ and hence
\[
\Pr(Y\ge t)\ \le\ \frac{\sigma^2}{\sigma^2+t^2}.
\]
The left-tail bound follows by applying the same argument to $-Y$. Both the inequalities in Eq.~(13) of the main text can be derived from the above theorem. See, refs~\cite{BoucheronLugosiMassart2013, Feller1970, GrimmettStirzaker2001, Klenke2014} for more details.

\section{Code description}\label{sec:code_description}
In this section, we describe the code used for the Monte-Carlo optimization. In particular, we focus on the time-evolution, which is computationally the most expensive part of the code. 

We work in the basis $\mathcal B = \{\ket{n_1,\cdots,n_L}: \sum_i n_i =N\}$ for $N$ bosons in $L$ modes. The dimension is $d=\binom{N+L-1}{L-1}$ and therefore quantum states $\ket{\psi}$ are $1-$d complex arrays with length $d$. During the $k-$th step in the evolution under Eq.~(16) of the main text, the propagation $\ket{\psi}\mapsto e^{-i\Delta t H_1^{(k)}}\ket{\psi}$ can de done efficiently, in $d$ steps because $H_1^{(k)}$ is diagonal in this basis. The propagations $\ket{\psi}\mapsto e^{-i\Delta t H_2^{(k)}}\ket{\psi}$ and $\ket{\psi}\mapsto e^{-i\Delta t H_3^{(k)}}\ket{\psi}$ are more challenging computationally.

Note that both $H_2^{(k)}$ and $H_3^{(k)}$ are sums of commuting terms and each term represents a hop from site $i$ to site $i+1$. We define a function that efficiently computes the propagation $\ket{\psi}\mapsto e^{-i\Delta t J_{i,i+1}(\hat{a}_i^{\dagger}\hat{a}_{i+1} +\hat{a}_{i+1}^{\dagger}\hat{a}_{i}) }\ket{\psi}$. Both $\ket{\psi}\mapsto e^{-i\Delta t H_2^{(k)}}\ket{\psi}$ and $\ket{\psi}\mapsto e^{-i\Delta t H_3^{(k)}}\ket{\psi}$ are a product of such operations.

We describe how the map $\ket{\psi} \mapsto e^{-i\Delta t J_{i,i+1}(\hat{a}_i^{\dagger}\hat{a}_{i+1} +\hat{a}_{i+1}^{\dagger}\hat{a}_{i}) }\ket{\psi}$ can be computed efficiently. Let us consider the case $i=1$ --- the arguments generalize to arbitrary $i$. The key observation is that the operator $(\hat{a}_1^{\dagger}\hat{a}_{2} +\hat{a}_{2}^{\dagger}\hat{a}_{1})$ has a block structure. It maps $\ket{n_1, n_2, \cdots, n_L}$ to $\ket{n_1', n_2', n_3\cdots, n_L}$, where the integers $n_3,\cdots, n_L$ remain the same and $n_1+ n_2= n_1' +n_2'$. We define vector spaces $V_r = \text{span}\{\ket{n_1, n_2, \cdots, n_L}: \ \  n_1+n_2=r\}$ for $r=0, \cdots, N$, which are invariant under $(\hat{a}_1^{\dagger}\hat{a}_{2} +\hat{a}_{2}^{\dagger}\hat{a}_{1})$. This suggests the following algorithm to compute the map $\ket{\psi} \mapsto e^{-i\Delta t J_{1,2}(\hat{a}_1^{\dagger}\hat{a}_{2} +\hat{a}_{2}^{\dagger}\hat{a}_{1}) }\ket{\psi}$:
\begin{itemize}
    \item[i.] Divide the vector $\psi$ into \textit{subvectors} (i.e., projections) $\ket{\psi_0}, \cdots, \ket{\psi_{N}}$ where $\ket{\psi_r}$ is the projection of $\ket{\psi}$ onto the space $V_r$.
    \item[ii.] Reshape $\ket{\psi_r}$ into an $(r+1)\times \binom{N-r+L-3}{N-r}$ matrix, where the $j-$th row consists of coefficients of the basis elements of the form $\ket{j, r-j, n_3, \cdots, n_L}$, for $j=0, \cdots, r$.
    \item[iii.] Compute the $(r+1)\times(r+1)$ unitary $U_r$ projection of $e^{-i\Delta t J_{1,2}(\hat{a}_1^{\dagger}\hat{a}_{2} +\hat{a}_{2}^{\dagger}\hat{a}_{1}) }$ by exponentiating the $(r+1)\times (r+1)$ matrix  $(\hat{a}_1^{\dagger}\hat{a}_{2} +\hat{a}_{2}^{\dagger}\hat{a}_{1})$ obtained by considering $r$ atoms in two modes --- $\hat{a}_1^{\dagger}, \hat{a}_2^{\dagger}$.
    \item[iv.] Left multiply $\ket{\psi_r}$, after reshaping into the $(r+1)\times \binom{N-r+L-3}{N-r}$ matrix by $U_r$. Reshaping this back into a 1D vector gives the projection of the final vector $e^{-i\Delta t J_{1,2}(\hat{a}_1^{\dagger}\hat{a}_{2} +\hat{a}_{2}^{\dagger}\hat{a}_{1}) }\ket{\psi}$ onto $V_r$. 
\end{itemize}

With this technique, $\ket{\psi}\mapsto e^{-i\Delta t H_2^{(k)}}\ket{\psi}$ and $\ket{\psi}\mapsto e^{-i\Delta t H_3^{(k)}}\ket{\psi}$ can both be computed in $\mathcal{O}(N^2\times L\times d)$ steps. In the code, we define a function  \texttt{pair\_hop} that includes the above steps. The function \texttt{pair\_hop}(i,\texttt{B}) organizes a bosonic occupation-number basis into blocks keyed by the \emph{pair occupancy} on two adjacent modes. Given a basis list \texttt{B} for \(N\) bosons in \(L\) modes—each element is a tuple \((n_0,\dots,n_{L-1})\) with \(\sum_j n_j = N\)—and an index \(i\) with \(0 \le i < L-1\), the function returns a list \([X_0,\dots,X_N]\). For each \(r\), the matrix \(X_r\) has shape \((r+1, K_r)\) and contains indices into \texttt{B}. Row \(s\) of \(X_r\) fixes \(n_i = s\) and \(n_{i+1} = r-s\) (hence \(n_i + n_{i+1} = r\)), while the columns enumerate all lexicographically ordered assignments of the remaining \(L-2\) occupations that sum to \(N-r\). Consequently, \(\texttt{B}[X_r[s,c]]\) equals the basis state with that pair \((n_i, n_{i+1})\) and the \(c\)-th configuration of the other modes.

Implementation-wise, the routine infers \(N\) and \(L\) from \texttt{B[0]}, builds a dictionary mapping occupation tuples to their positions in \texttt{B}, and scans \texttt{B} once to group states by \(r = n_i + n_{i+1}\) while collecting the “other” coordinates (with positions \(i,i+1\) removed). Sorting these “others” sets fixes a consistent column order across rows in each \(X_r\). It then reconstructs full tuples by inserting \((s, r-s)\) back into each column’s “others” tuple and fills \(X_r\) with their indices via the dictionary. The number of columns is \(K_r = \binom{N - r + L - 3}{N - r}\) (the stars-and-bars count for distributing \(N-r\) bosons over \(L-2\) modes), and edge cases are handled naturally (e.g., \(L=2\) yields only \(X_N\) with a single column)

\section{Quantum Optimal Control for Small Systems}\label{sec:QOC}
\begin{figure}[t]
    \centering
    \includegraphics[width=0.99\linewidth]{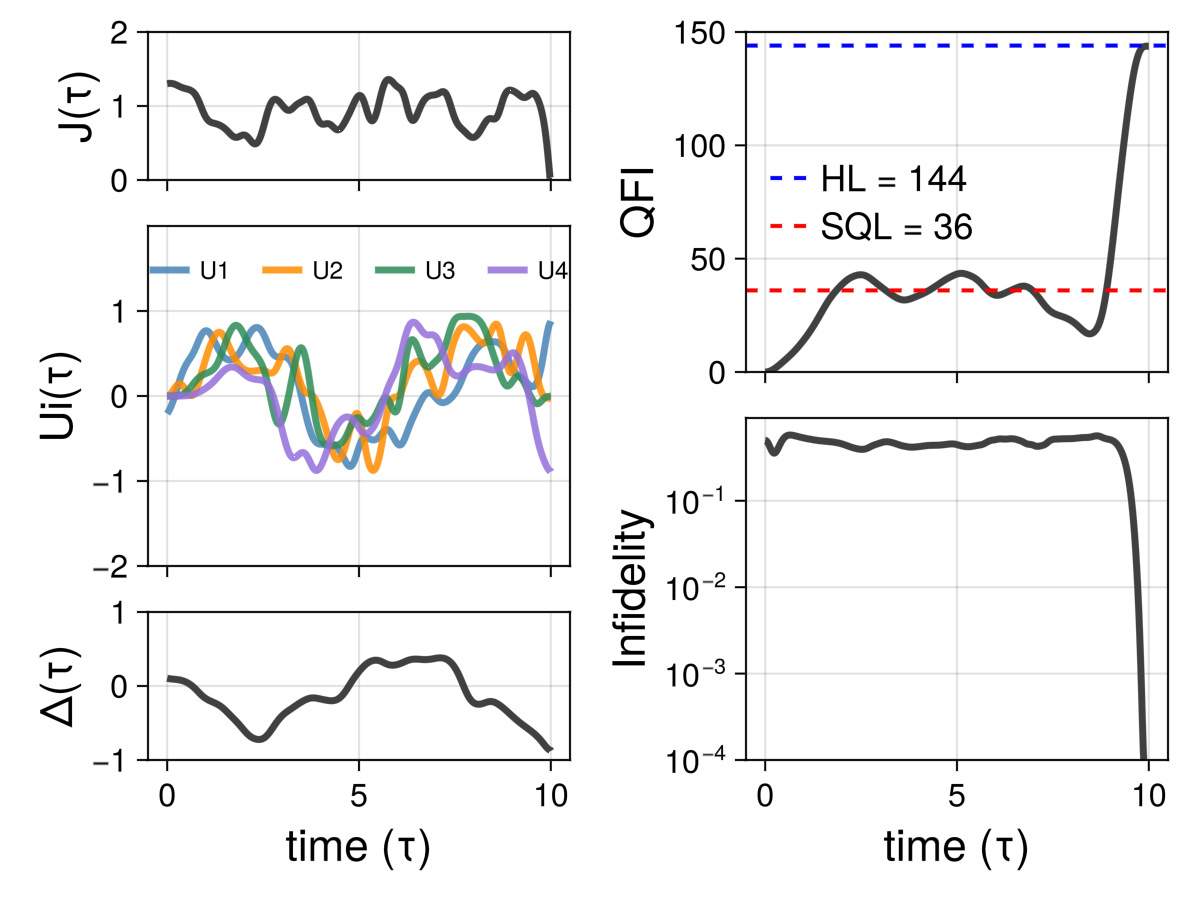}
    \caption{\textbf{Left:} The pulse sequence that steers the system from a Fock state $\ket{4000}$ to a GHZ-like state. \textbf{Right Top:} The QFI of the state $\ket{\psi(t)}$ evolves in time. The initial QFI is zero since all bosons are in site 1 at $t=0$, and the final QFI is very close to the HL since the final state is a GHZ-like state. \textbf{Right Bottom:} The infidelity between the quantum state $\ket{\psi(t)}$ and the GHZ-like state. Note that here we optimize the parameters $J(t)$, $U_i(t)$ and $\Delta(t)$ are explicitly smooth in contrast to Eq.~(10) in the main text, resulting in the smooth behavior of the QFI.}
    \label{fig_44_fock2GHZ}
\end{figure}
Since the QFI can be maximized by a GHZ-like state Eq.~\eqref{eq:GHZ} in the main text, we can approach the ultimate precision (HL) by performing a state preparation: our initial state is a Fock state $\ket{N0\dots 0}$, where all $N$ bosons are in the first site, and the target state is a GHZ-like state $\ket{\psi_{\text{GHZ}}}$. This state transfer problem can be solved by QOC, and the quantum optimal control problem (Q-OCP) can be written as   
\begin{subequations}\label{eq:QOCP_GHZ}
\begin{eqnarray}
\displaystyle \max~~ && \lvert \braket{\psi(T)|\psi_{\text{GHZ}}}\rvert ^2\\
\textrm{s.t.}~~&& \ket{\dot{\psi}(t)}\!=\!H(t)\ket{\psi(t)}\!, \ket{\psi(0)}\!=\!\ket{N0\dots 0}
\end{eqnarray}
\end{subequations}
where we wish to maximize the fidelity between the final quantum state $\ket{\psi(T)}$ and the target state $\ket{\psi_{\text{GHZ}}}$, the quantum state $\ket{\psi(t)}$ is governed by the time-dependent Hamiltonian
\begin{equation}\label{eq:continuous_H}
    \begin{gathered}
    H(t) = J(t)\sum^{L-1}_{i=1}(\hat{a}^{\dagger}_{i}\hat{a}_{i+1} + \hat{a}^{\dagger}_{i+1}\hat{a}_{i}) + \sum^{L}_{i=1}U_{i}(t)\hat{n}_{i}(\hat{n}_i - 1)\\ +  \Delta(t)\sum_{i=1}^{L}i\hat{n}_i
    \end{gathered}
\end{equation}

Note that the parameters $J(t)$, $U_i(t)$ and $\Delta(t)$ are explicitly smooth, in contrast to Eq.~\eqref{eq:H} in the main text. The well-defined Q-OCP Eq.~\eqref{eq:QOCP_GHZ} can be solved using Q-PRONTO~\cite{Shao2022,Shao2024}, which is a Newton-based QOC solver that can achieve quadratic convergence and find the optimal solution directly in function space.
\begin{figure}
    \centering
    \includegraphics[width=0.99\linewidth]{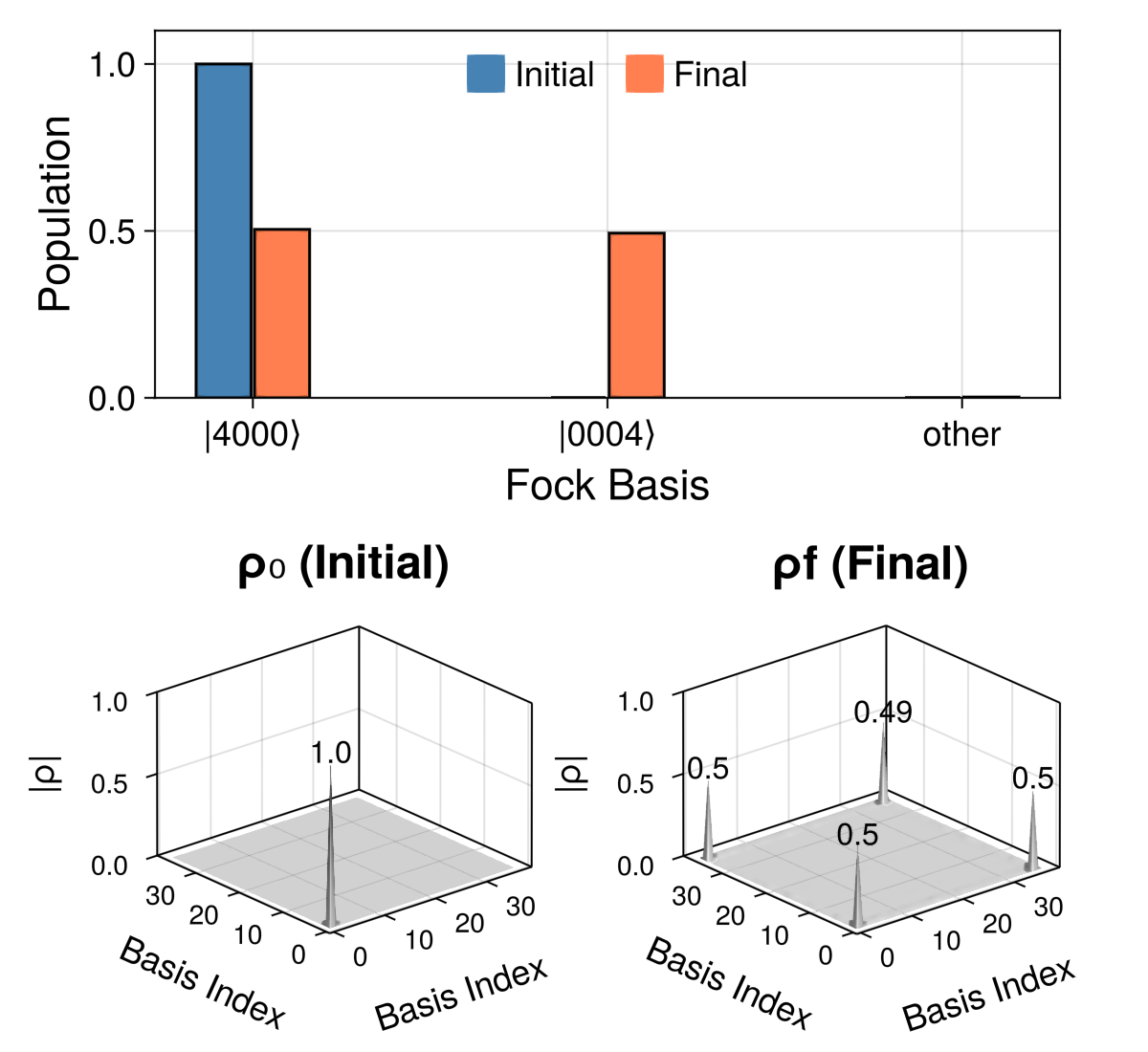}
    \caption{\textbf{Top:} The initial and final population of two Fock states, $\ket{4000}$ and $\ket{0004}$. Since GHZ-like state is a superposition between the two Fock states, half of the population is in $\ket{4000}$ while the other half in $\ket{0004}$. \textbf{Bottom:} Amplitude of the initial and final density matrices $\rho_0 = \ket{\psi(0)}\bra{\psi(0)}, \rho_f = \ket{\psi(T)}\bra{\psi(T)}$. The Fock basis index ranges from 1 to 35, starting with Fock state $\ket{4000}$ and ending with $\ket{0004}$.}
    \label{fig_44noon_pop}
\end{figure}

Consider a small system with 4 bosons ($N=4$) and 4 lattice sites ($L=4$). We will show that we are able to steer the system from an initial Fock state $\ket{4000}$ to the target GHZ-like state $(\ket{4000}+e^{i\phi}\ket{0004})/\sqrt{2}$ using Q-PRONTO. Figure \ref{fig_44_fock2GHZ} shows the control parameters $J(t), U_i(t), \Delta(t)$ that steer the system from the initial state $\ket{4000}$ to the target state $\ket{\psi_{\text{GHZ}}}$, the QFI time evolution and the infidelity between $\ket{\psi(t)}$ and $\ket{\psi_{\text{GHZ}}}$ over time. The time horizon $T=10 \tau$, where $\tau = J^{-1}$ is the unit of time. The QFI initially starts at 0, exceeds the $\text{SQL} =(L-1)^2N = 36$, and finally reaches 143.73, close to the $\text{HL}=(L-1)^2N^2 = 144$. Figure \ref{fig_44noon_pop} shows the initial and final population in Fock basis (first index is $\ket{4000}$ and last index is $\ket{0004}$), and the amplitude of the initial and final density matrices $|\rho|=|\ket{\psi}\bra{\psi}|$. Since the initial state is a Fock state $\ket{4000}$, we see a single spike with amplitude 1; the final state is close to a GHZ-like state, and thus we see four spikes at the corners with the amplitude 0.5. While we showcase that the HL can be approached for small system, it is hard to implement QOC to large systems: the dimension of the Hilbert space (Eq.~\eqref{eq:dim} in the main text) grows exponentially, QOC suffers from the curse of dimensionality and its performance becomes poor.  

\end{document}